\documentclass[twocolumn]{aastex62}

\usepackage{textcomp}
\usepackage{bm,color}
\usepackage{verbatim}
\usepackage{amsmath}
\usepackage{ulem}
\usepackage{hyperref}  % <- new
\usepackage{booktabs}
\usepackage{bigstrut}
\usepackage{xspace}
\usepackage{multirow}
\usepackage{amsmath}
\usepackage{graphicx}
\usepackage{threeparttable}

\newcommand{\spt}{{\sc spt}}
\newcommand{\sptpol}{{\sc SPTpol}}
\newcommand{\sptsz}{{\sc SPT-SZ}\xspace}
\newcommand{\planckTT}{{\sc PlanckTT}\xspace}
\newcommand{\lowP}{{\sc lowP}\xspace}

\newcommand{\lcdm}{$\Lambda$CDM\xspace}

\newcommand{\nver}{\hat{\mathbf{n}}}
\newcommand{\Dd}{\mathbf{d}}
\newcommand*\Bell{\ensuremath{\boldsymbol\ell}}

\newcommand*\BTheta{\ensuremath{\boldsymbol\Theta}}

\newcommand{\vanish}[1]{}
\newcommand{\LCDM}{\ensuremath{\Lambda {\rm CDM}}}
\newcommand{\planck}{\textit{Planck}\xspace}
\newcommand{\sqdeg}{$\deg^2$}

\newcommand{\sigmaOmSPTpol}{\ensuremath{0.593 \pm 0.025}}
\newcommand{\sigmaOmPlanck}{\ensuremath{0.590 \pm 0.020}}
\newcommand{\sigmaOmPlanckSPTpol}{\ensuremath{0.587 \pm 0.015}}

\newcommand{\sigmaPlanck}{\ensuremath{0.829 \pm 0.015}}
\newcommand{\sigmaPlanckLens}{\ensuremath{0.820 \pm 0.010}}
\newcommand{\sigmaSPTpolLens}{\ensuremath{0.816 \pm 0.012}}
\newcommand{\OmegamSPTpolLensBAO}{\ensuremath{0.368^{+0.032}_{-0.037}}}
\newcommand{\sigmaSPTpolLensBAO}{\ensuremath{0.779 \pm 0.023}}
\newcommand{\AphiphiSPTpol}{\ensuremath{0.890^{+0.057}_{-0.066}}}
\newcommand{\AphiphiPlanck}{\ensuremath{0.970\pm 0.039}}
\newcommand{\AlensAphiphiSPTpol}{\ensuremath{0.995\pm0.090}}
\newcommand{\AlensAphiphiSPTpolSPTpol}{\ensuremath{1.036\pm0.136}}
\newcommand{\AlensAphiphiPlanck}{\ensuremath{1.076\pm 0.063}}
\newcommand{\omegaKPlanck}{\ensuremath{-0.043^{+0.028}_{-0.016}}}

\newcommand{\omegaKPlanckLens}{\ensuremath{-0.0084^{+0.0093}_{-0.0076}}}
\newcommand{\omegaKSPTpolLens}{\ensuremath{-0.0099^{+0.013}_{-0.0084}}}
\newcommand{\omegaKSPTpolLensBAO}{\ensuremath{-0.0007 \pm 0.0025}}
\newcommand{\mnuPlanckLensBAO}{\ensuremath{0.22}}
\newcommand{\mnuSPTpolLensBAO}{\ensuremath{0.23}}
\newcommand{\mnuSPTpolTEEESPTpolLensBAO}{\ensuremath{0.42}}
\newcommand{\mnuPlanckLensBAOAlens}{\ensuremath{0.39}}
\newcommand{\mnuSPTpolLensBAOAlens}{\ensuremath{0.45}}
\newcommand{\mnuSPTpolTEEESPTpolLensBAOAlens}{\ensuremath{0.62}}

\newcommand{\beq}{\begin{equation}}
\newcommand{\eeq}{\end{equation}}
\newcommand{\bea}{\begin{eqnarray}}
\newcommand{\eea}{\end{eqnarray}}
\newenvironment{rcases}
  {\left.\begin{aligned}}
  {\end{aligned}\right\rbrace}
  
\begin{document}
%%%%%%%%%%%%%%%%%%%%% Title, etc. %%%%%%%%%%%%%%%%%%%%%
\title{Constraints on Cosmological Parameters from the 500 \sqdeg{} \sptpol{} Lensing Power Spectrum}

\shortauthors{F.~Bianchini, W.~L.~K.~Wu, et al.}
\author[0000-0003-4847-3483]{F.~Bianchini} \affiliation{School of Physics, University of Melbourne, Parkville, VIC 3010, Australia}
\author[0000-0001-5411-6920]{W.~L.~K.~Wu} \affiliation{Kavli Institute for Cosmological Physics, University of Chicago, 5640 South Ellis Avenue, Chicago, IL, USA 60637}
\author{P.~A.~R.~Ade} \affiliation{Cardiff University, Cardiff CF10 3XQ, United Kingdom}
\author{A.~J.~Anderson} \affiliation{Fermi National Accelerator Laboratory, MS209, P.O. Box 500, Batavia, IL 60510}
\author{J.~E.~Austermann} \affiliation{NIST Quantum Devices Group, 325 Broadway Mailcode 817.03, Boulder, CO, USA 80305} \affiliation{Department of Physics, University of Colorado, Boulder, CO, USA 80309}
\author{J.~S.~Avva} \affiliation{Department of Physics, University of California, Berkeley, CA, USA 94720}
\author{J.~A.~Beall} \affiliation{NIST Quantum Devices Group, 325 Broadway Mailcode 817.03, Boulder, CO, USA 80305}
\author{A.~N.~Bender} \affiliation{High Energy Physics Division, Argonne National Laboratory, 9700 S. Cass Avenue, Argonne, IL, USA 60439} \affiliation{Kavli Institute for Cosmological Physics, University of Chicago, 5640 South Ellis Avenue, Chicago, IL, USA 60637}
\author[0000-0002-5108-6823]{B.~A.~Benson} \affiliation{Fermi National Accelerator Laboratory, MS209, P.O. Box 500, Batavia, IL 60510} \affiliation{Kavli Institute for Cosmological Physics, University of Chicago, 5640 South Ellis Avenue, Chicago, IL, USA 60637} \affiliation{Department of Astronomy and Astrophysics, University of Chicago, 5640 South Ellis Avenue, Chicago, IL, USA 60637}
\author[0000-0001-7665-5079]{L.~E.~Bleem} \affiliation{High Energy Physics Division, Argonne National Laboratory, 9700 S. Cass Avenue, Argonne, IL, USA 60439} \affiliation{Kavli Institute for Cosmological Physics, University of Chicago, 5640 South Ellis Avenue, Chicago, IL, USA 60637}
\author{J.~E.~Carlstrom} \affiliation{Kavli Institute for Cosmological Physics, University of Chicago, 5640 South Ellis Avenue, Chicago, IL, USA 60637} \affiliation{Department of Physics, University of Chicago, 5640 South Ellis Avenue, Chicago, IL, USA 60637} \affiliation{High Energy Physics Division, Argonne National Laboratory, 9700 S. Cass Avenue, Argonne, IL, USA 60439} \affiliation{Department of Astronomy and Astrophysics, University of Chicago, 5640 South Ellis Avenue, Chicago, IL, USA 60637} \affiliation{Enrico Fermi Institute, University of Chicago, 5640 South Ellis Avenue, Chicago, IL, USA 60637}
\author{C.~L.~Chang} \affiliation{Kavli Institute for Cosmological Physics, University of Chicago, 5640 South Ellis Avenue, Chicago, IL, USA 60637} \affiliation{High Energy Physics Division, Argonne National Laboratory, 9700 S. Cass Avenue, Argonne, IL, USA 60439} \affiliation{Department of Astronomy and Astrophysics, University of Chicago, 5640 South Ellis Avenue, Chicago, IL, USA 60637}
\author{P.~Chaubal} \affiliation{School of Physics, University of Melbourne, Parkville, VIC 3010, Australia}
\author{H.~C.~Chiang} \affiliation{Department of Physics, McGill University, 3600 Rue University, Montreal, Quebec H3A 2T8, Canada} \affiliation{School of Mathematics, Statistics \& Computer Science, University of KwaZulu-Natal, Durban, South Africa}
\author{R.~Citron} \affiliation{University of Chicago, 5640 South Ellis Avenue, Chicago, IL, USA 60637}
\author{C.~Corbett~Moran} \affiliation{TAPIR, Walter Burke Institute for Theoretical Physics, California Institute of Technology, 1200 E California Blvd, Pasadena, CA, USA 91125}
\author[0000-0001-9000-5013]{T.~M.~Crawford} \affiliation{Kavli Institute for Cosmological Physics, University of Chicago, 5640 South Ellis Avenue, Chicago, IL, USA 60637} \affiliation{Department of Astronomy and Astrophysics, University of Chicago, 5640 South Ellis Avenue, Chicago, IL, USA 60637}
\author{A.~T.~Crites} \affiliation{Kavli Institute for Cosmological Physics, University of Chicago, 5640 South Ellis Avenue, Chicago, IL, USA 60637} \affiliation{Department of Astronomy and Astrophysics, University of Chicago, 5640 South Ellis Avenue, Chicago, IL, USA 60637} \affiliation{California Institute of Technology, MS 249-17, 1216 E. California Blvd., Pasadena, CA, USA 91125}
\author{T.~de~Haan} \affiliation{Department of Physics, University of California, Berkeley, CA, USA 94720} \affiliation{Physics Division, Lawrence Berkeley National Laboratory, Berkeley, CA, USA 94720}
\author{M.~A.~Dobbs} \affiliation{Department of Physics, McGill University, 3600 Rue University, Montreal, Quebec H3A 2T8, Canada} \affiliation{Canadian Institute for Advanced Research, CIFAR Program in Gravity and the Extreme Universe, Toronto, ON, M5G 1Z8, Canada}
\author{W.~Everett} \affiliation{Department of Astrophysical and Planetary Sciences, University of Colorado, Boulder, CO, USA 80309}
\author{J.~Gallicchio} \affiliation{Kavli Institute for Cosmological Physics, University of Chicago, 5640 South Ellis Avenue, Chicago, IL, USA 60637} \affiliation{Harvey Mudd College, 301 Platt Blvd., Claremont, CA 91711}
\author{E.~M.~George} \affiliation{European Southern Observatory, Karl-Schwarzschild-Str. 2, 85748 Garching bei M\"{u}nchen, Germany} \affiliation{Department of Physics, University of California, Berkeley, CA, USA 94720}
\author{A.~Gilbert} \affiliation{Department of Physics, McGill University, 3600 Rue University, Montreal, Quebec H3A 2T8, Canada}
\author{N.~Gupta} \affiliation{School of Physics, University of Melbourne, Parkville, VIC 3010, Australia}
\author{N.~W.~Halverson} \affiliation{Department of Astrophysical and Planetary Sciences, University of Colorado, Boulder, CO, USA 80309} \affiliation{Department of Physics, University of Colorado, Boulder, CO, USA 80309}
\author{N.~Harrington} \affiliation{Department of Physics, University of California, Berkeley, CA, USA 94720}
\author{J.~W.~Henning} \affiliation{High Energy Physics Division, Argonne National Laboratory, 9700 S. Cass Avenue, Argonne, IL, USA 60439} \affiliation{Kavli Institute for Cosmological Physics, University of Chicago, 5640 South Ellis Avenue, Chicago, IL, USA 60637}
\author{G.~C.~Hilton} \affiliation{NIST Quantum Devices Group, 325 Broadway Mailcode 817.03, Boulder, CO, USA 80305}
\author[0000-0002-0463-6394]{G.~P.~Holder} \affiliation{Astronomy Department, University of Illinois at Urbana-Champaign, 1002 W. Green Street, Urbana, IL 61801, USA} \affiliation{Department of Physics, University of Illinois Urbana-Champaign, 1110 W. Green Street, Urbana, IL 61801, USA} \affiliation{Canadian Institute for Advanced Research, CIFAR Program in Gravity and the Extreme Universe, Toronto, ON, M5G 1Z8, Canada}
\author{W.~L.~Holzapfel} \affiliation{Department of Physics, University of California, Berkeley, CA, USA 94720}
\author{J.~D.~Hrubes} \affiliation{University of Chicago, 5640 South Ellis Avenue, Chicago, IL, USA 60637}
\author{N.~Huang} \affiliation{Department of Physics, University of California, Berkeley, CA, USA 94720}
\author{J.~Hubmayr} \affiliation{NIST Quantum Devices Group, 325 Broadway Mailcode 817.03, Boulder, CO, USA 80305}
\author{K.~D.~Irwin} \affiliation{SLAC National Accelerator Laboratory, 2575 Sand Hill Road, Menlo Park, CA 94025} \affiliation{Dept. of Physics, Stanford University, 382 Via Pueblo Mall, Stanford, CA 94305}
\author{L.~Knox} \affiliation{Department of Physics, University of California, One Shields Avenue, Davis, CA, USA 95616}
\author{A.~T.~Lee} \affiliation{Department of Physics, University of California, Berkeley, CA, USA 94720} \affiliation{Physics Division, Lawrence Berkeley National Laboratory, Berkeley, CA, USA 94720}
\author{D.~Li} \affiliation{NIST Quantum Devices Group, 325 Broadway Mailcode 817.03, Boulder, CO, USA 80305} \affiliation{SLAC National Accelerator Laboratory, 2575 Sand Hill Road, Menlo Park, CA 94025}
\author{A.~Lowitz} \affiliation{Department of Astronomy and Astrophysics, University of Chicago, 5640 South Ellis Avenue, Chicago, IL, USA 60637}
\author{A.~Manzotti} \affiliation{Kavli Institute for Cosmological Physics, University of Chicago, 5640 South Ellis Avenue, Chicago, IL, USA 60637} \affiliation{Institut d'Astrophysique de Paris, 98 bis boulevard Arago, 75014 Paris, France}
\author{J.~J.~McMahon} \affiliation{Department of Physics, University of Michigan, 450 Church Street, Ann  Arbor, MI, USA 48109}
\author{S.~S.~Meyer} \affiliation{Kavli Institute for Cosmological Physics, University of Chicago, 5640 South Ellis Avenue, Chicago, IL, USA 60637} \affiliation{Department of Physics, University of Chicago, 5640 South Ellis Avenue, Chicago, IL, USA 60637} \affiliation{Department of Astronomy and Astrophysics, University of Chicago, 5640 South Ellis Avenue, Chicago, IL, USA 60637} \affiliation{Enrico Fermi Institute, University of Chicago, 5640 South Ellis Avenue, Chicago, IL, USA 60637}
\author{M.~Millea} \affiliation{Department of Physics, University of California, Berkeley, CA, USA 94720}
\author{L.~M.~Mocanu} \affiliation{Kavli Institute for Cosmological Physics, University of Chicago, 5640 South Ellis Avenue, Chicago, IL, USA 60637} \affiliation{Department of Astronomy and Astrophysics, University of Chicago, 5640 South Ellis Avenue, Chicago, IL, USA 60637}
\author{J.~Montgomery} \affiliation{Department of Physics, McGill University, 3600 Rue University, Montreal, Quebec H3A 2T8, Canada}
\author{A.~Nadolski} \affiliation{Astronomy Department, University of Illinois at Urbana-Champaign, 1002 W. Green Street, Urbana, IL 61801, USA} \affiliation{Department of Physics, University of Illinois Urbana-Champaign, 1110 W. Green Street, Urbana, IL 61801, USA}
\author{T.~Natoli} \affiliation{Department of Astronomy and Astrophysics, University of Chicago, 5640 South Ellis Avenue, Chicago, IL, USA 60637} \affiliation{Kavli Institute for Cosmological Physics, University of Chicago, 5640 South Ellis Avenue, Chicago, IL, USA 60637} \affiliation{Dunlap Institute for Astronomy \& Astrophysics, University of Toronto, 50 St George St, Toronto, ON, M5S 3H4, Canada}
\author{J.~P.~Nibarger} \affiliation{NIST Quantum Devices Group, 325 Broadway Mailcode 817.03, Boulder, CO, USA 80305}
\author{G.~Noble} \affiliation{Department of Physics, McGill University, 3600 Rue University, Montreal, Quebec H3A 2T8, Canada}
\author{V.~Novosad} \affiliation{Materials Sciences Division, Argonne National Laboratory, 9700 S. Cass Avenue, Argonne, IL, USA 60439}
\author{Y.~Omori} \affiliation{Kavli Institute for Particle Astrophysics and Cosmology, Stanford University, 452 Lomita Mall, Stanford, CA 94305}
\author{S.~Padin} \affiliation{Kavli Institute for Cosmological Physics, University of Chicago, 5640 South Ellis Avenue, Chicago, IL, USA 60637} \affiliation{Department of Astronomy and Astrophysics, University of Chicago, 5640 South Ellis Avenue, Chicago, IL, USA 60637} \affiliation{California Institute of Technology, MS 249-17, 1216 E. California Blvd., Pasadena, CA, USA 91125}
\author{S.~Patil} \affiliation{School of Physics, University of Melbourne, Parkville, VIC 3010, Australia}
\author{C.~Pryke} \affiliation{School of Physics and Astronomy, University of Minnesota, 116 Church Street S.E. Minneapolis, MN, USA 55455}
\author[0000-0003-2226-9169]{C.~L.~Reichardt} \affiliation{School of Physics, University of Melbourne, Parkville, VIC 3010, Australia}
\author{J.~E.~Ruhl} \affiliation{Physics Department, Center for Education and Research in Cosmology and Astrophysics, Case Western Reserve University, Cleveland, OH, USA 44106}
\author{B.~R.~Saliwanchik} \affiliation{Physics Department, Center for Education and Research in Cosmology and Astrophysics, Case Western Reserve University, Cleveland, OH, USA 44106} \affiliation{Department of Physics, Yale University, P.O. Box 208120, New Haven, CT 06520-8120}
\author{J.T.~Sayre} \affiliation{Department of Astrophysical and Planetary Sciences, University of Colorado, Boulder, CO, USA 80309} \affiliation{Department of Physics, University of Colorado, Boulder, CO, USA 80309}
\author{K.~K.~Schaffer} \affiliation{Kavli Institute for Cosmological Physics, University of Chicago, 5640 South Ellis Avenue, Chicago, IL, USA 60637} \affiliation{Enrico Fermi Institute, University of Chicago, 5640 South Ellis Avenue, Chicago, IL, USA 60637} \affiliation{Liberal Arts Department, School of the Art Institute of Chicago, 112 S Michigan Ave, Chicago, IL, USA 60603}
\author{C.~Sievers} \affiliation{University of Chicago, 5640 South Ellis Avenue, Chicago, IL, USA 60637}
\author{G.~Simard} \affiliation{Department of Physics, McGill University, 3600 Rue University, Montreal, Quebec H3A 2T8, Canada}
\author{G.~Smecher} \affiliation{Department of Physics, McGill University, 3600 Rue University, Montreal, Quebec H3A 2T8, Canada} \affiliation{Three-Speed Logic, Inc., Vancouver, B.C., V6A 2J8, Canada}
\author{A.~A.~Stark} \affiliation{Harvard-Smithsonian Center for Astrophysics, 60 Garden Street, Cambridge, MA, USA 02138}
\author{K.~T.~Story} \affiliation{Kavli Institute for Particle Astrophysics and Cosmology, Stanford University, 452 Lomita Mall, Stanford, CA 94305} \affiliation{Dept. of Physics, Stanford University, 382 Via Pueblo Mall, Stanford, CA 94305}
\author{C.~Tucker} \affiliation{Cardiff University, Cardiff CF10 3XQ, United Kingdom}
\author{K.~Vanderlinde} \affiliation{Dunlap Institute for Astronomy \& Astrophysics, University of Toronto, 50 St George St, Toronto, ON, M5S 3H4, Canada} \affiliation{Department of Astronomy \& Astrophysics, University of Toronto, 50 St George St, Toronto, ON, M5S 3H4, Canada}
\author{T.~Veach} \affiliation{Department of Astronomy, University of Maryland College Park, MD, USA 20742}
\author{J.~D.~Vieira} \affiliation{Astronomy Department, University of Illinois at Urbana-Champaign, 1002 W. Green Street, Urbana, IL 61801, USA} \affiliation{Department of Physics, University of Illinois Urbana-Champaign, 1110 W. Green Street, Urbana, IL 61801, USA}
\author{G.~Wang} \affiliation{High Energy Physics Division, Argonne National Laboratory, 9700 S. Cass Avenue, Argonne, IL, USA 60439}
\author[0000-0002-3157-0407]{N.~Whitehorn} \affiliation{Department of Physics and Astronomy, University of California, Los Angeles, CA, USA 90095}
\author{V.~Yefremenko} \affiliation{High Energy Physics Division, Argonne National Laboratory, 9700 S. Cass Avenue, Argonne, IL, USA 60439} %for aas submission

\begin{abstract}
We present cosmological constraints based on the cosmic microwave background (CMB) lensing potential power spectrum measurement from the recent 500\,\sqdeg{} \sptpol{} survey, the most precise CMB lensing measurement from the ground to date. 
We fit a flat \lcdm{} model to the reconstructed lensing power spectrum alone and in addition with other data sets: baryon acoustic oscillations (BAO) as well as primary CMB spectra from \planck and \sptpol{}.
The cosmological constraints based on \sptpol{} and \planck lensing band powers are in good agreement when analysed alone and in combination with \planck full-sky primary CMB  data.
With weak priors on the baryon density and other parameters, the \sptpol{} CMB lensing data alone provide a 4\% constraint on $\sigma_8\Omega_m^{0.25} = \sigmaOmSPTpol$.
Jointly fitting with BAO data, we find $\sigma_8=\sigmaSPTpolLensBAO$, $\Omega_m = \OmegamSPTpolLensBAO$, and $H_0 = 72.0^{+2.1}_{-2.5}\,\text{km}\,\text{s}^{-1}\,\text{Mpc}^{-1} $, up to $2\,\sigma$ away from the central values preferred by \planck lensing + BAO. However, we recover good agreement between \sptpol{} and \planck when restricting the analysis to similar scales.
We also consider single-parameter extensions to the flat \lcdm model. The \sptpol{} lensing spectrum constrains the spatial curvature to be $\Omega_K = \omegaKSPTpolLensBAO$ and the sum of the neutrino masses to be $\sum m_{\nu} < \mnuSPTpolLensBAO$ eV at  95\% C.L. (with \planck primary CMB and BAO data), in good agreement with the \planck lensing results. 
With the differences in the $S/N$ of the lensing modes and the angular scales covered in the lensing spectra, this analysis represents an important independent check on the full-sky \planck lensing measurement. 
\end{abstract}


%%%%%%%%
% INTRO
%%%%%%%%
\section{Introduction}

Measurements of the gravitational lensing of the cosmic microwave background (CMB) by large-scale structure provide a unique observational probe of the geometry of the universe and the growth of structure at high redshifts. 
As light travels from the last-scattering surface to us, the paths of the photons are bent by the gravitational potential of matter. 
These deflections are related to the gradient of the gravitational potential and can be used to reconstruct the gravitational potential integrated along the line of sight \citep{lewis06}. 
Gravitational lensing of the CMB also provides a powerful tool for constraining neutrino masses, since massive neutrinos suppress structure growth \citep[e.g.,][]{lesgourgues06b, abazajian15b}. 
CMB lensing has been measured by a number of experiments using both temperature and polarization data \citep[e.g.,][]{das11, vanengelen12, planck13-17, polarbear2014a, story14, planck15-15, bicep2keck16, sherwin16, omori17, planck18-8,wu19}. 
The most precise lensing amplitude measurement to date comes from \planck, which measures the overall lensing amplitude at 40\,$\sigma$ \citep{planck18-8}.

Intriguingly, there is a modest level of discordance between the primary CMB power spectra from \planck and other cosmological probes within the \LCDM {} model. 
Relevant to the case of lensing, the amplitude of density fluctuations $\sigma_8$ deduced from galaxy cluster counts and cosmic shear measurements is slightly 
lower than the value suggested by primary CMB \planck data \citep[e.g.,][]{hildebrandt17,abbott18,bocquet19,hikage19,zubeldia19,joudaki19}.
Tensions within the \planck dataset are also emerging, for example the amount of lensing inferred from the smoothing of the acoustic peaks in the \planck CMB power spectra 
is larger than the one directly measured through the CMB lensing potential power spectrum \citep{planck18-6}. 
Whether these tensions have their origins in unaccounted for systematics, new physics, or are simply statistical fluctuations is not yet clear and more detailed analyses are needed in order to shed light on these discrepancies.

One way to probe if these apparent tensions are caused by systematics is to use measurements from independent experiments. 
In this work, we infer cosmological parameters using the high-$S/N$ lensing power spectrum measurement from the 500\,\sqdeg{} \sptpol{} survey \citep[][hereafter W19]{wu19}, currently the most precise CMB lensing measurement from the ground. 
While measured over only 1\% of the sky, the \sptpol{} lensing amplitude uncertainty is only twice as large as the uncertainty of the \planck lensing measurement from 67\% of the sky. 
Thus the \sptpol\ lensing power spectrum provides a chance to test for  consistency between CMB lensing measurements performed over different fractions of the sky and angular scales.
In particular, the \sptpol\ lensing power spectrum complements the \planck{} lensing measurements by extending the measurement to smaller angular scales. 
If the two lensing measurements are consistent, their combination has the potential to improve our cosmological model constraints.

In this work, we explore the cosmological implications of the high-significance measurement of the lensing angular power spectrum from W19. 
Within the \lcdm model, we begin by comparing cosmological parameters inferred from the \sptpol{} lensing measurements against those from \planck and optical surveys. We then contrast parameters from lensing measurements and primary CMB measurements.
After that, we look at what these lensing measurements tell us about the curvature of the universe and the sum of the neutrino masses, as well as other model extensions using a suite of Monte Carlo Markov chains. 
As in W19, we take the best-fit \LCDM{} model for the \planckTT + \lowP + {\sc lensing}  dataset in \citet{planck15-13}  to be our fiducial model.

%organization
This paper is organized as follows. 
We outline the principles of CMB lensing and how the lensing potential can be reconstructed in Sec.~\ref{sec:lensing_reco}. 
In Sec.~\ref{sec:constraints}, we explore cosmological parameters constraints from the lensing data in different models. 
Finally, we draw our conclusions in Sec.~\ref{sec:conclusions}.

%%%%%%%%
% LENSING RECO FRAMEWORK
%%%%%%%%
\section{Lensing reconstruction framework}
\label{sec:lensing_reco}
In this section we briefly review the physics of CMB lensing, sketch the lensing reconstruction pipeline steps in the context of \sptpol, and describe the CMB lensing likelihood 
modelling. For a thorough description of the \sptpol\ lensing reconstruction analysis, we refer the reader to W19.

\subsection{Basics of CMB lensing}
\label{sec:lensing_basics}
During their journey from the last-scattering surface to us, CMB photons are deflected by the gradients of gravitational potentials associated with the large-scale 
structure (LSS) \citep{blanchard87,bernardeau97,lewis06}. As a result, the unlensed CMB temperature $T(\nver)$ and polarization  $[Q\pm i U](\nver)$ anisotropies 
are remapped according to:

\beq
\tilde{X}(\nver) = X(\nver + \Dd(\nver)),
\eeq
where $X(\nver)$ denotes either the temperature or polarization fluctuations in a given direction of the sky $\nver$, and the tilde indicates the lensed quantities. At lowest order, the deflection field $\Dd(\nver)$ can be written as the angular gradient of the Weyl gravitational potential $\Psi$ projected along the line-of-sight, $\Dd(\nver) = \nabla \phi(\nver)$, where we have introduced the CMB lensing potential $\phi$:

\beq
\label{eq:cmb_lens_phi}
\phi(\nver) = -2\int_0^{\chi_{\rm CMB}} d\chi \frac{f_K(\chi_{\rm CMB}-\chi)}{f_K(\chi_{\rm CMB})f_K(\chi)} \Psi(\chi\nver,\eta_0-\chi).
\eeq
Here, $\chi$ is the comoving distance (with $\chi_{\rm CMB}\approx$ 14000 Mpc denoting the distance to the last scattering surface), $f_K$ is the angular-diameter distance, 
and $\eta_0-\chi$ is the conformal time. The divergence of the deflection field gives the lensing convergence $\kappa = -\frac{1}{2}\nabla^2\phi$, that quantifies the amount of 
local (de)magnification of the CMB fluctuations.

Eq.~\ref{eq:cmb_lens_phi} tells us that CMB lensing probes both the geometry and the growth of structure of the universe and as such, precise measurements of its power 
spectrum can break the geometrical degeneracy affecting the primary CMB \citep{stompor99} and tighten constraints on the sum of neutrino masses $\sum m_{\nu}$ as well as on 
the amplitude of density fluctuations $\sigma_8$ \citep{smith09}.

\subsection{Lensing extraction with quadratic estimators}
\label{sec:lensing_QE}
Lensing correlates previously independent CMB temperature and polarization modes between different angular scales on the sky.
This lensing-induced correlation is the basis for lensing quadratic estimators, which reconstruct the lensing potential $\phi$ by examining the correlation between
CMB Fourier modes \citep{zaldarriaga99,hu02a}. 

The formally optimal estimator (at lowest order in 
$\phi$) has the following form

\beq
\label{eq:lens_est}
\bar{\phi}^{XY}_{\bf L} = \int d^2 \Bell \, W_{\Bell,\Bell-{\bf L}}^{XY} \bar{X}_{\Bell}\bar{Y}_{\Bell-{\bf L}}^*,
\eeq
where $\bar{X}$ and $\bar{Y}$ are the filtered $T$, $E$, or $B$ fields, and $W_{\Bell,\Bell-{\bf L}}^{XY}$ is a weight function (unique to the $XY$ pair, see \citealt{hu02a} for the 
exact expressions).
We recall that in the W19 lensing analysis only CMB modes with $|\Bell_x|>100$ and $|\Bell | < 3000$ are used, to account for the impact of time-stream filtering and mitigate foreground contamination. The input CMB maps are filtered with an inverse-variance (C$^{-1}$) filter to down-weight noisy modes and to increase the sensitivity to lensing. In addition, 
the unlensed CMB spectra in the weights $W_{\Bell,\Bell-{\bf L}}^{XY}$ are replaced with the lensed ones to cancel higher-order biases \citep{hanson11}. 

The lensing potential $\bar{\phi}^{XY}_{\bf L}$ measured with Eq.~\ref{eq:lens_est} is a biased estimate of the true lensing potential $\phi^{XY}_{\bf L}$:

\beq
\bar{\phi}^{XY}_{\bf L} = \mathcal{R}_{\bf L}^{XY}\phi^{XY}_{\bf L},
\eeq
where $\mathcal{R}_{\bf L}^{XY}$ is a response function that normalizes the estimator. As discussed in W19, the response function adopted in this analysis is first calculated 
analytically and then corrected perturbatively with simulations, $\mathcal{R}_{\bf L}^{XY} = \mathcal{R}_{\bf L}^{XY, \rm Analytic}\mathcal{R}_{\bf L}^{XY, \rm MC}$ . To illustrate 
the cosmological dependence of this response function, which will be relevant to calculate the corrections to the lensing likelihood in Sec.~\ref{sec:lensing_like}, we explicitly 
write down the analytical response function in the case of an isotropic filter:

\beq
\mathcal{R}_{\bf L}^{XY, \rm Analytic} = \int d^2 \Bell \, W_{\Bell,\Bell-{\bf L}}^{XY} W_{\Bell,\Bell-{\bf L}}^{XY} \mathcal{F}^X_{\Bell} \mathcal{F}^{Y}_{\Bell-{\bf L}},
\eeq
where $\mathcal{F}^X_{\ell} = \left (C_{\ell}^{XX} + N_{\ell}^{XX}\right)^{-1}$. Note that both the filters $\mathcal{F}_{\ell}$ and the weight functions $W_{\ell,\ell - L}$ are calculated assuming a fiducial cosmology. 

Anisotropic features such as inhomogeneous noise and coupling of modes due to masking introduce spurious signals that mimic the effects of lensing.
To circumvent this, we remove a mean-field correction $\bar{\phi}_{\bf L}^{XY,\rm MF}$ estimated by averaging $\bar{\phi}$ reconstructed from many input lensed CMB simulations. The final estimate of the lensing potential is

\beq
\hat{\phi}^{XY}_{\bf L} = \frac{1}{\mathcal{R}_{\bf L}^{XY}} \left (\bar{\phi}_{\bf L}^{XY} - \bar{\phi}_{\bf L}^{XY,\rm MF} \right).
\eeq
Finally, the different lensing estimators $XY \in \{ TT,TE,\allowbreak TB,EE,EB, ET, BT, BE \}$ are combined into a minimum-variance (MV) estimate using

\beq
\hat{\phi}^{\rm MV}_{\bf L} = \frac{1}{\mathcal{R}_{\bf L}^{\rm MC}} \frac{ \sum_{XY} \bar{\phi}_{\bf L}^{XY} - \bar{\phi}_{\bf L}^{XY,\rm MF} }{\sum_{XY} \mathcal{R}_{\bf L}^{XY, \rm Analytic} } .
\eeq

\subsection{Power spectrum estimation}
\label{sec:lensing_spectra}
Cosmological inference is carried out by comparing the measured CMB lensing power spectrum to the theoretical expectations over the parameter space. 
After obtaining the unbiased lensing potential $\hat{\phi}$, the raw CMB lensing potential power spectrum $C_L^{\hat{\phi}^{XY}\hat{\phi}^{ZW}}$ is measured by forming cross-
spectra of $\hat{\phi}^{XY}_{\bf L}$ and $\hat{\phi}^{ZW}_{\bf L}$. The resulting power spectrum is a biased estimate of the true CMB lensing power spectrum. In W19, 
four sources of biases are corrected for \citep{hu02a,kesden03,hanson11}:

\beq
\hat{C}^{\phi\phi}_L = f_{\rm PS} \left[ C_L^{\hat{\phi}\hat{\phi}} - N_L^{(0),RD} - N_L^{(1)} \right] - \Delta C_L^{\phi\phi, \rm FG}.
\eeq
$N_L^{(0)}$ is the disconnected (Gaussian) bias term that arises from chance correlations in the CMB, noise, and foregrounds. We estimate it with the realization-dependent 
method described in \cite{namikawa13} that reduces the covariance between lensing band powers and eliminates the dependency on the fiducial cosmology at linear order. 
Secondary contractions of the connected 4-point function source an additional bias term, known as $N^{(1)}$, that depends linearly on the true CMB lensing potential power 
spectrum and hence, on the cosmological parameters. In the flat-sky limit, and assuming an isotropic filtering, it can be evaluated as 
\citep{kesden03,planck15-15}

\begin{equation}
\begin{split}
N^{(1)}_L  =& \frac{1}{\mathcal{R}^{XZ}_L\mathcal{R}^{CD}_L} \int \frac{d^2\Bell_1}{(2\pi)^2} \frac{d^2\Bell'_1}{(2\pi)^2} \\
&\times \mathcal{F}^X_{\ell_1} \, \mathcal{F}^Z_{\ell_2} \, \mathcal{F}^C_{\ell'_1} \, \mathcal{F}^D_{\ell'_2}\,  W^{ZD}_{-\Bell_2,\Bell'_2} \, W^{ZD}_{-\Bell_2,\Bell'_2} \\
&\times \Big[ C^{\phi\phi}_{|\Bell_1-\Bell'_1|} W^{XC}_{-\Bell_1,\Bell'_1} W^{ZD}_{-\Bell_2,\Bell'_2} \\
&+ C^{\phi\phi}_{|\Bell_1-\Bell'_2|} W^{XD}_{-\Bell_1,\Bell'_2} W^{ZC}_{-\Bell_2,\Bell'_1} \Big], 
\end{split}
\end{equation}
where $\Bell_1 + \Bell_2 = \Bell'_1+\Bell'_2={\bf L}$. In W19, this bias term is estimated using simulations, as done in \citet{story14}.

Foreground emission can introduce biases in the reconstructed lensing map and the lensing power spectrum, especially if correlated with the LSS. In particular, thermal Sunyaev-Zel'dovich (tSZ) effect and cosmic infrared background (CIB) emission can leak into the reconstructed lensing map correlating with the lensing potential. 
In addition, tSZ and CIB have trispectra that can leak into the CMB lensing spectrum through the 4-point function of the temperature map.
Adopting the bias estimates from \citet{vanengelen14a}, we remove a foreground bias term $\Delta C_L^{\phi\phi, \rm FG}$ from the temperature components of the MV spectrum that include tSZ trispectra, CIB trispectra, tSZ$^2-\phi$ and CIB$^2-\phi$ contributions.

Higher order biases, like $N^{(2)}_L$ are cancelled by the use of lensed CMB spectra in the lensing weights $W_{\ell,\ell - L}$ \citep{hanson11}, while biases induced by the non-Gaussianity of the LSS or by the post-Born corrections are negligible at the current $S/N$ \citep[e.g.,][]{pratten2016,bohm2016,beck2018}.

Any non-idealities not captured by the lensing reconstruction analysis might result in discrepancies between the input theory and the recovered amplitude in simulations. 
We refer to this residual bias as ``Monte Carlo bias" and, in our analysis, we find that the main source of this bias is higher-order coupling generated by the presence of the point-source mask, and we rescale the measured lensing power spectrum by a multiplicative correction $f_{\rm PS}$ of order 5\% to account for this effect.

Finally, the lensing bandpower covariance is estimated using $N_s = 400$ Monte Carlo sky realizations that have been fully processed through the lensing analysis pipeline. 
Specifically, the input CMB maps are passed through a mock observing pipeline that uses the pointing information to produce mock time-ordered data from these simulated skies for each SPT detector, filters those data in the same fashion as the real data, and generates maps using the inverse-noise weights from the real data.

\begin{figure}
	\includegraphics[width=\columnwidth]{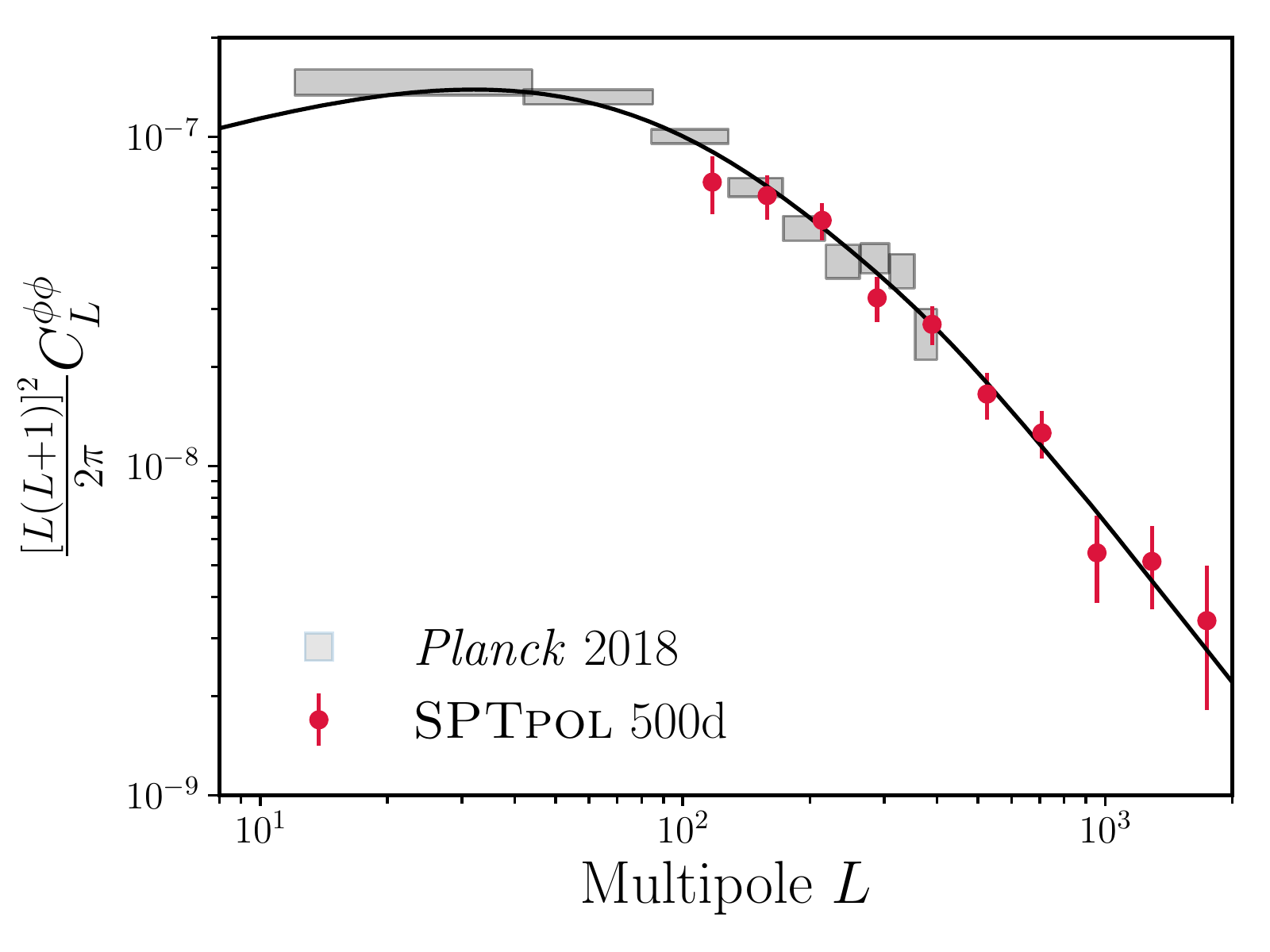}
    \caption{CMB lensing potential power spectrum measurements from \sptpol{} 500d \citep[][red circles]{wu19} and from \planck 2018 \citep[][grey boxes]{planck18-8}. The black solid line is the lensing spectrum from the best-fit \lcdm model to the 2015  \planckTT + \lowP + {\sc lensing}  dataset.}
    \label{fig:sptpol_clpp}
\end{figure}

The final CMB lensing band powers used for the cosmological analysis presented in this paper are shown by the red circles in Fig.~\ref{fig:sptpol_clpp}, together with the 2018 \planck $C_L^{\phi\phi}$ measurement.

\subsection{CMB lensing likelihood}
\label{sec:lensing_like}
We approximate the lensing log-likelihood as Gaussian in the band powers of the estimated lensing power spectrum:

\beq
\label{eq:lens_like}
\begin{split}
-2 \ln{\mathcal{L}_{\phi}}&(\BTheta) = \\
\sum_{ij}& \left[ \hat{C}_{L^i_{\rm b}}^{\phi\phi} - C_{L^i_{\rm b}}^{\phi\phi,\rm th} (\BTheta) \right] \mathbb{C}^{-1}_{L^i_{\rm b}L^j_{\rm b}}  \left[ \hat{C}_{L^j_{\rm b}}^{\phi\phi} - C_{L^j_{\rm b}}^{\phi\phi,\rm th} (\BTheta) \right],
\end{split}
\eeq
where $C_{L^i_{\rm b}}^{\phi\phi,\rm th} (\BTheta)$ is the binned theory spectrum at the position $\BTheta$ in the parameter space. In Eq.~\ref{eq:lens_like} we ignore the
correlations between the 2- and 4-point functions since these have been shown to be negligible at current sensitivities \citep{schmittfull13,peloton17}. 
In practice, this means that when combining the CMB power spectra from \textit{Planck} with \sptpol\ lensing, we simply multiply their respective likelihoods.
The covariance matrix $\mathbb{C}^{-1}_{L^i_{\rm b}L^j_{\rm b}} $ is calculated using Monte Carlo simulations and includes small off-diagonal elements.
When inverting the covariance matrix, we neglect the correction from \citet{hartlap06} as this is only a $\approx 2-3\%$ effect. 
For completeness, we also point out that we do not inflate our covariance matrix by a $\approx 4\%$ factor to account for the Monte Carlo uncertainties arising from using a finite number of simulations to estimate the mean-field and the noise biases \citep{planck18-8}.

\begin{figure}
	\includegraphics[width=\columnwidth]{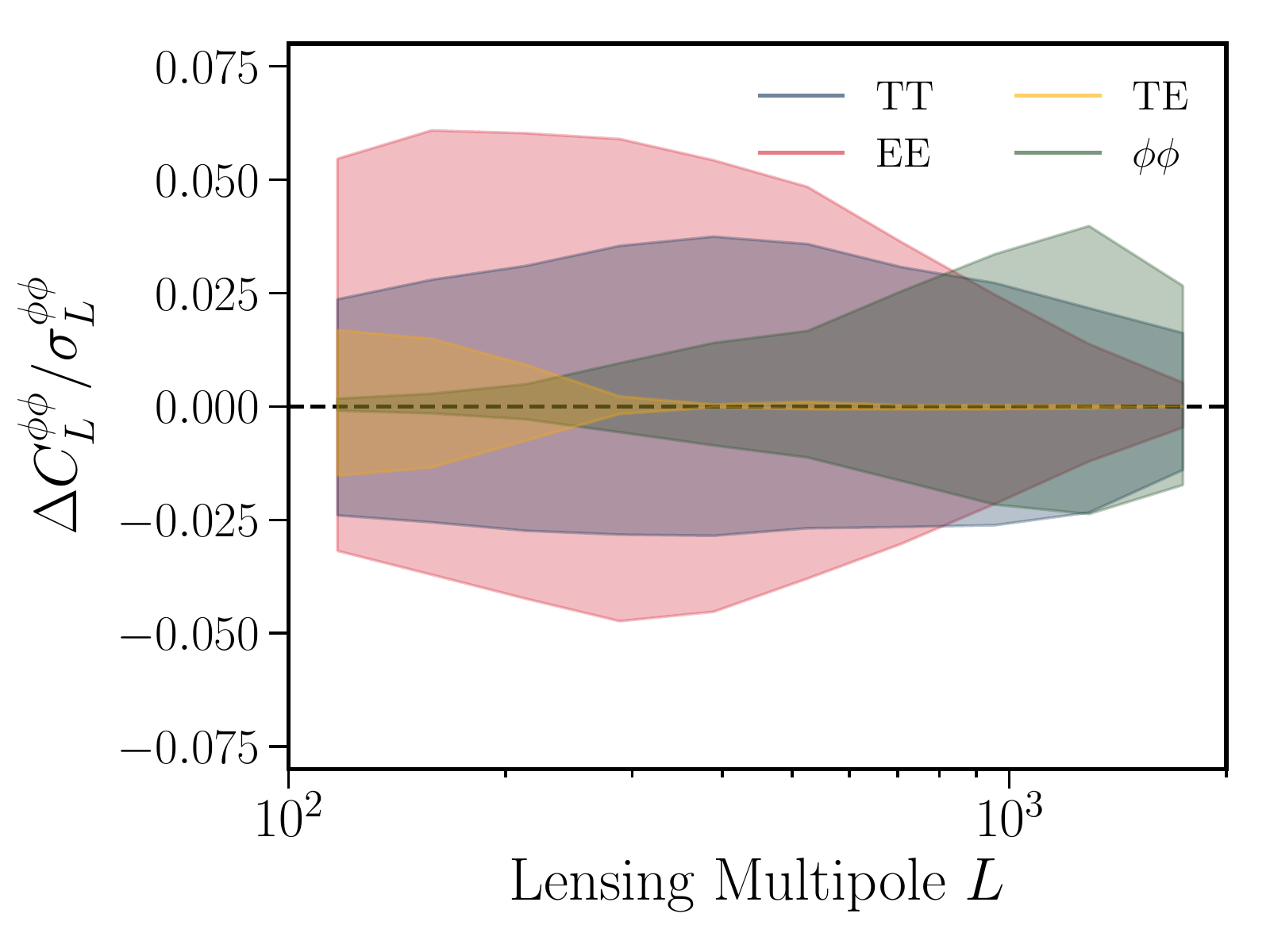}
    \caption{Ratio of the lensing likelihood corrections, $\Delta C_L^{\phi\phi} = M^a_{LL'} \left( C_{L'}^a |_{\BTheta} - C_{L'}^a |_{\rm fid} \right)$, to the \sptpol{} lensing
    band powers uncertainties $\sigma_L^{\phi\phi}$ for the CMB power spectra $TT/TE/EE$ and $\phi\phi$ corrections (blue, yellow, red, and cyan bands respectively). 
    The different bands contain the 68\% of the correction distributions evaluated for points $\BTheta$ in the parameter space randomly drawn from the \planckTT + \lowP + {\sc lensing} cosmology chains from \citet{planck15-13}.}
    \label{fig:corrections_sptpol}
\end{figure}

The fiducial cosmology assumed in the lensing reconstruction affects the estimated lensing band powers.
The underlying cosmological parameters do not only enter Eq.~\ref{eq:lens_like} through the theoretical lensing power spectrum $C_L^{\phi\phi,\rm th} (\BTheta)$, but also
indirectly through the calculation of the response functions $\mathcal{R}_L$ and the $N^{(1)}$ bias. For a given pair of quadratic estimators $x$ and $y$, the 
corrected theory lensing power spectrum can be written as

\beq
C_L^{\phi\phi,\rm th}|_{\BTheta}   = \frac{( \mathcal{R}_L^{x} \mathcal{R}_L^{y} )|_{\BTheta}}{( \mathcal{R}_L^{x} \mathcal{R}_L^{y} ) |_{\rm fid}} C_L^{\phi\phi}\big |_{\BTheta} + N_L^{(1)xy}\big |_{\BTheta} - N_L^{(1)xy}\big |_{\rm fid}.
\eeq
Evaluating these quantities at each point in the parameter space is computationally unfeasible, therefore we follow the approach of \citet{planck15-15,sherwin16,simard18} and perturbatively correct the theory spectrum for changes due to the parameter deviations from the fiducial cosmology. For such small deviations, we can Taylor-expand the response function and the $N^{(1)}_L$ bias around the fiducial cosmology and obtain:\footnote{We neglect the dependence of the $N^{(1)}_L$ bias on the CMB power spectra. Also note that we use isotropic approximations to model both the response function and the $N^{(1)}_L$ bias.}

\beq
\begin{split}
C_L^{\phi\phi,\rm th}|_{\BTheta}  &\approx \frac{\partial \ln( \mathcal{R}_L^{x} \mathcal{R}_L^{y} )}{\partial C_{\ell'}^j}  \left( C_{\ell'}^j |_{\BTheta} - C_{\ell'}^j |_{\rm fid} \right)C_{\ell'}^{\phi\phi} |_{\rm fid}  \\
&+ \frac{\partial N^{(1)xy}_L}{\partial C_{L'}^{\phi\phi}}\left( C_{L'}^{\phi\phi} |_{\BTheta} - C_{L'}^{\phi\phi} |_{\rm fid} \right)\\
&= C_L^{\phi\phi}\big|_{\BTheta} + M^a_{LL'} \left( C_{L'}^a |_{\BTheta} - C_{L'}^a |_{\rm fid} \right)\\
\end{split}
\eeq
where summation over repeated indices is implied, $j$ sums over the CMB power spectra $TT$, $TE$, and $EE$, while $a$ also sums over $\phi\phi$ in addition to $TT$, $TE$, and $EE$. The correction matrices $M^a_{LL'}$ can then be pre-computed for the fiducial model and binned.
We make use of the publicly available \texttt{quicklens}\footnote{Available at \url{https://github.com/dhanson/quicklens}.} and \texttt{lensingbiases}\footnote{Available at \url{https://github.com/JulienPeloton/lensingbiases}} packages to calculate the derivative with respect to the response function and the $N^{(1)}_L$ bias, respectively. Finally, for the MV CMB lensing power spectrum, we coadd the different $xy$ linear corrections according to the MV weights, as done for the real data. 
To give a sense of the magnitude and the spectral dependence of the different lensing corrections, 
we show their breakdown in Fig.~\ref{fig:corrections_sptpol},
evaluated for 100 points in the parameter space randomly drawn from the \planck chains corresponding to our fiducial cosmology.

%

%%%%%%%%
% CONSTRAINTS
%%%%%%%%
\section{Cosmological parameter constraints}
\label{sec:constraints}
In this section we investigate the constraining power of the \sptpol\ lensing dataset on cosmology and compare to \textit{Planck} lensing constraints.

\subsection{Cosmological inference framework}
\label{sec:cosmo_inference_framework}
Our reference cosmological model is a spatially flat \lcdm model with purely adiabatic scalar primordial fluctuations and a single family of massive neutrinos with total mass $\sum m_{\nu}=60$ meV. This baseline model is described by a set of six parameters: the physical baryon density $\Omega_b h^2$, the physical cold dark matter density $\Omega_c h^2$, the 
(approximated) angular size of the sound horizon at recombination $\theta_{\rm MC}$, the optical depth at reionization $\tau$, the amplitude $A_s$ and spectral index $n_s$ of primordial scalar fluctuations calculated at a pivot scale of $k=0.05$ Mpc$^{-1}$. We will also quote parameters derived from these six parameters, such as the total matter density $\Omega_m$, the Hubble constant $H_0$, and the amplitude of the matter power spectrum expressed in terms of $\sigma_8$, the rms density fluctuations within a sphere of radius 8 $h^{-1}$ Mpc.
We calculate the lensed CMB and CMB lensing potential power spectra with the \texttt{camb}\footnote{Available at \url{https://camb.info} (August 2017 version). The small-scale nonlinear matter power spectrum and its effect on the CMB lensing quantities are calculated with the \texttt{HMcode} of \citet{mead15}. 
The effect of the non-linear matter power spectrum modelling uncertainties on the estimated cosmological parameters has been shown to be negligible by \citet{planck18-8}, who compared constraints obtained adopting both the \citet{takahashi12} and \citet{mead15} versions of the \texttt{halofit} model. The impact of modelling differences is found to be negligible even when considering the full multipole range $8 \le L \le 2048$.} Boltzmann code \citep{lewis00}, while the parameter posteriors are sampled with the Markov Chain Monte Carlo (MCMC) \texttt{CosmoMC}\footnote{Available at \url{https://cosmologist.info/cosmomc/}  (July 2018 version) } code \citep{lewis02b}.
 
In the following we combine the likelihoods associated to five different datasets: i) the 2015 \planckTT and \lowP primary CMB likelihoods \citep{planck15-11}; ii) the 2018 \textit{Planck} CMB lensing likelihood \citep{planck18-8}; iii) the \sptpol\ CMB lensing likelihood\footnote{Details on how to install and use the SPTpol CMB lensing likelihood and dataset are available at \url{https://pole.uchicago.edu/public/data/lensing19/}.}; iv) the \sptpol{} $TEEE$ likelihood \citep{henning18}; v)  baryonic acoustic oscillation (BAO) likelihoods from BOSS DR12, SDSS MGS, and 6dFGS galaxy surveys data \citep{beutler11,ross15,alam17}.

We do not use the latest \planck 2018 primary CMB data because the \planck 2018 likelihoods were only publicly released when the analysis and preparation of this manuscript were near completion. Since the main aim of this work is to compare the constraining power of the \sptpol{} and \planck lensing datasets, this does not represent an issue as long as we combine them with the same primary CMB datasets. For completeness, we recall that the main differences between the 2015 and 2018 \planck releases are an improved processing of the low-$\ell$ HFI polarization data and the inclusion of polarization corrections in the high-$\ell$ likelihood (not used here), whose principal effect is to lower the central value and tighten the uncertainty by a factor of 2 on $\tau$. Consequently, the $A_se^{-2\tau}$ degeneracy causes a 1$\sigma$ decrease of $\ln(10^{10}A_s)$ and a $\approx 0.5\sigma$ increase of $\Omega_c h^2$.

\begin{table}[t]
\centering
\caption{Summary of the priors imposed on each cosmological parameter in this work, when considering either lensing-only datasets or also including primary CMB measurements. Parameters that are fixed are reported by a single number. $\mathcal{U}(a,b)$ denotes a uniform distribution between $[a,b]$, while $\mathcal{N}(\mu,\sigma^2)$ indicates a Gaussian distribution with mean $\mu$ and variance $\sigma^2$. }
\label{tab:priors}
\begin{tabular}{c|c|c}
Parameter               & Lensing only                  & Lensing +  CMB      \\ 
\hline
\hline

$\Omega_b h^2$          & $\mathcal{N}(0.0222,0.0005^2)$ & $\mathcal{U}(0.005,0.1)$     \\
$\Omega_c h^2$          & $\mathcal{U}(0.001,0.99)$      & $\mathcal{U}(0.001,0.99)$    \\
$H_0$ [km/s/Mpc]                & $\mathcal{U}(40,100)$          & $\mathcal{U}(40,100)$        \\
$\tau$                  & 0.055                          & $\mathcal{U}(0.01,0.8)$      \\
$n_s$                   & $\mathcal{N}(0.96,0.02^2)$     & $\mathcal{U}(0.8,1.2)$       \\
$\ln (10^{10}A_s)$        & $\mathcal{U}(1.61,3.91)$       & $\mathcal{U}(1.61,3.91)$     \\
$\sum m_{\nu}$ {[}eV{]} & 0.06                           & 0.06 or $\mathcal{U}(0,5)$   \\
$\Omega_K$              & 0                              & 0 or $\mathcal{U}(-0.3,0.3)$ \\
$A_L$                   & 1                              & 1 or $\mathcal{U}(0,10)$     \\
$A_L^{\phi\phi}$        & 1                              & 1 or $\mathcal{U}(0,10)$    
\end{tabular}%
\end{table}

\subsection{Constraints from CMB lensing alone}
\label{sec:constraints_lensing_only}
We start by showing the constraints on the baseline \lcdm model using only CMB lensing measurements.
In particular, we focus on the amplitude of the matter power spectrum $\sigma_8$ and the total matter density $\Omega_m$.
When analyzing constraints from CMB lensing alone, we follow \citet{planck18-8} and adopt the weak priors shown in Tab.~\ref{tab:priors} to avoid marginalizing over unrealistic values of poorly constrained parameters. Specifically, we fix the optical depth to reionization to $\tau=0.055$ and  place Gaussian priors on the baryon density $\Omega_bh^2=0.0222 \pm 0.0005$, motivated by primordial deuterium abundance D/H  measurements in high-redshift metal-poor quasar absorption systems \citep{cooke2018} combined with big-bang nucleosynthesis predictions, and on the spectral index $n_s= 0.96 \pm 0.02$. 
Moreover, we fix the linear corrections to the response function to the fiducial cosmology, similar to \citet{planck15-15,sherwin16,simard18}. 

\begin{figure}
	\includegraphics[width=\columnwidth]{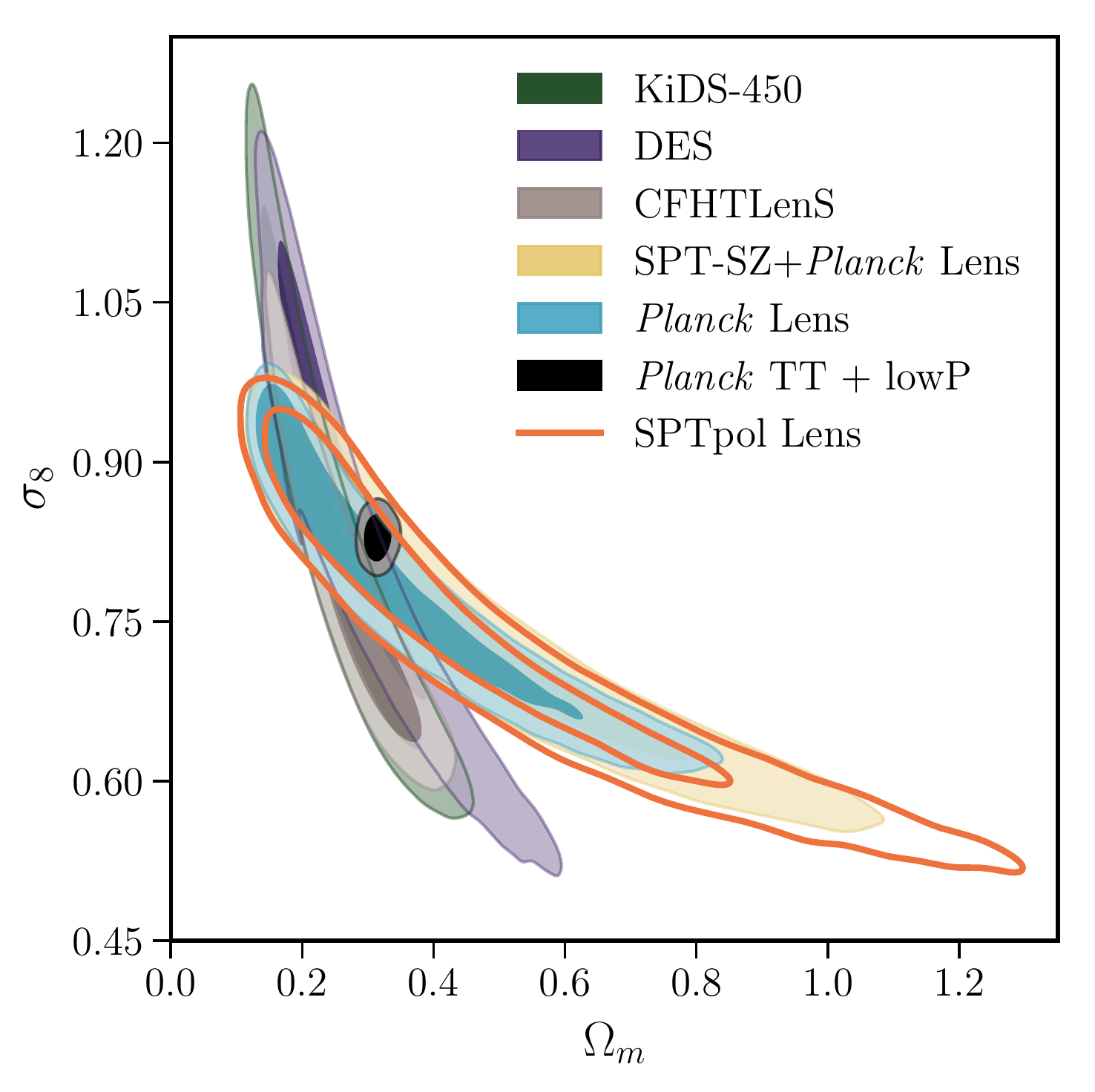}
    \caption{The constraints on $\sigma_8$ and $\Omega_m$ from CMB lensing (\sptpol{}, \planck, \sptsz+\planck) and optical lensing (KiDS-450, CFHTLenS, DES) surveys appear to be broadly consistent with each other. The different degeneracy direction between CMB and optical lensing surveys reflects their different redshift sensitivity to matter fluctuations. The independent high-redshift constraints from \planck primary CMB power spectra, shown as black contours, are also in agreement with the lower redshift CMB lensing measurements.}
    \label{fig:sigma8_omegam_lensonly}
\end{figure}

As shown in Fig.~\ref{fig:sigma8_omegam_lensonly}, the lensing-only constraints project a well-defined band in the $\Omega_m-\sigma_8$ plane:

\beq
\sigma_8\Omega_m^{0.25} = \sigmaOmSPTpol \quad (\text{\sptpol\ lensing only,}\,\, 68\%).
\eeq
This parameter combination is measured with a 4.2\% precision and is in excellent agreement with the \planck lensing-only value of $\sigma_8\Omega_m^{0.25} = \sigmaOmPlanck$ (3.4\% precision). 
For comparison, in Fig.~\ref{fig:sigma8_omegam_lensonly} we also show the constraints obtained by \citet{simard18} with the CMB lensing band powers from 2500 deg$^2$ observed by \sptsz + \planck \citep{omori17}, which are again consistent with the \sptpol\ ones and similar in extent.

Assuming that the \sptpol{} and \planck lensing measurements are independent, which is a safe assumption given the relatively small footprint overlap 
and the different sensitivity to CMB modes due to noise and resolution, we can further combine the datasets. 
The parameter that mostly benefits from the joint analysis is once again $\sigma_8\Omega_m^{0.25}$, which is constrained with an accuracy of $\approx 2.5\%$:

\beq
\begin{split}
\sigma_8\Omega_m^{0.25} = \sigmaOmPlanckSPTpol \quad (&\text{\planck + \sptpol\ }\\
&\text{lensing only,}\,\, 68\%).
\end{split}
\eeq
This corresponds to a factor of 1.33 improvement over \planck lensing-only statistical uncertainties. 

\subsection{Comparison with galaxy lensing}
\label{sec:galaxy_lensing}
While the focus of this analysis is interpreting the gravitational lensing measurements of the CMB, we are also able to compare our results with measurements of optical weak lensing, also known as cosmic shear  \citep[e.g.,][]{bartelmann01}. 
Cosmic shear is complementary to CMB lensing as it is sensitive to the evolution of gravitational potentials at lower redshift and is affected by different systematics. 
 
In Fig.~\ref{fig:sigma8_omegam_lensonly} we also show a compilation of recent cosmic shear constraints on $\Omega_m$ and $\sigma_8$ from the KiDS\footnote{We make use of the \texttt{kids450fiducial} chains available at \url{https://github.com/sjoudaki/kids450}.} \citep{hildebrandt17}, CFHTLenS\footnote{Results shown here are based on the \texttt{fiducialrun} chains available at \url{https://www.dropbox.com/s/lku48ron59nvc1m/centralchains.tar.gz?dl=0}.} \citep{joudaki17a}, and DES \citep{troxel18} optical surveys.
We follow the \citet{planck18-8} approach and use the first-year DES cosmic shear likelihood, data cuts, and nuisance parameters (and associated priors) described by \citet{troxel18}, but use the priors on cosmological parameters shown in Tab~\ref{tab:priors}. Furthermore, we consider a single minimal-mass neutrino eigenstate. 

As can be seen in Fig.~\ref{fig:sigma8_omegam_lensonly}, the statistical power of galaxy lensing 
constraints is comparable to that of  CMB lensing, but due to the much lower redshift distribution of the source galaxies, the degeneracy direction is different and approximately constrains $\sigma_8\Omega_m^{0.5} $. In fact, the parameter combination that is best constrained by cosmic shear measurements is $S_8 = \sigma_8 \sqrt{\Omega_m/0.3}$, which is measured at $5\%$ accuracy by DES lensing, $S_8 = 0.790^{+0.040}_{-0.029}$. 
This constraint is $\sim1.1\, \sigma$ lower than the value inferred from \planck 2018 primary CMB, $S_8 = 0.834 \pm 0.016$, and consistent with $S_8$ from \sptpol{} lensing, $S_8 = 0.86\pm 0.11$. 
A similar level of precision has been achieved by \citet{hikage19} who performed a tomographic analysis of the Hyper Suprime-Cam (HSC) survey first-year shear catalog and found $S_8 = 0.777^{+0.031}_{-0.034}$.\footnote{For consistency with the results based on CMB lensing and DES cosmic shear, here we quote the constraint obtained by fixing the sum of the neutrino masses to $\sum m_{\nu} = 0.06 $ eV.}

\subsection{Including BAO information}
\label{sec:lensing_bao}
We next consider the cosmological implication of adding BAO data to the lensing measurements.
In addition to $\sigma_8$ and $\Omega_m$, the lensing spectrum is sensitive to the expansion rate $H_0$ since it also constrains the parameter combination $\sigma_8\Omega_m^{0.25} ( \Omega_m h^2 )^{-0.37}$~\citep[e.g.][]{planck15-15}. To break the degeneracy, we include BAO measurements (and the Gaussian prior on $\Omega_b h^2$ from the column "Lensing only" in Tab.~\ref{tab:priors}), which allow the BAO measurements to constrain $H_0$ and $\Omega_m$.

\begin{figure}[t]
	\includegraphics[width=\columnwidth]{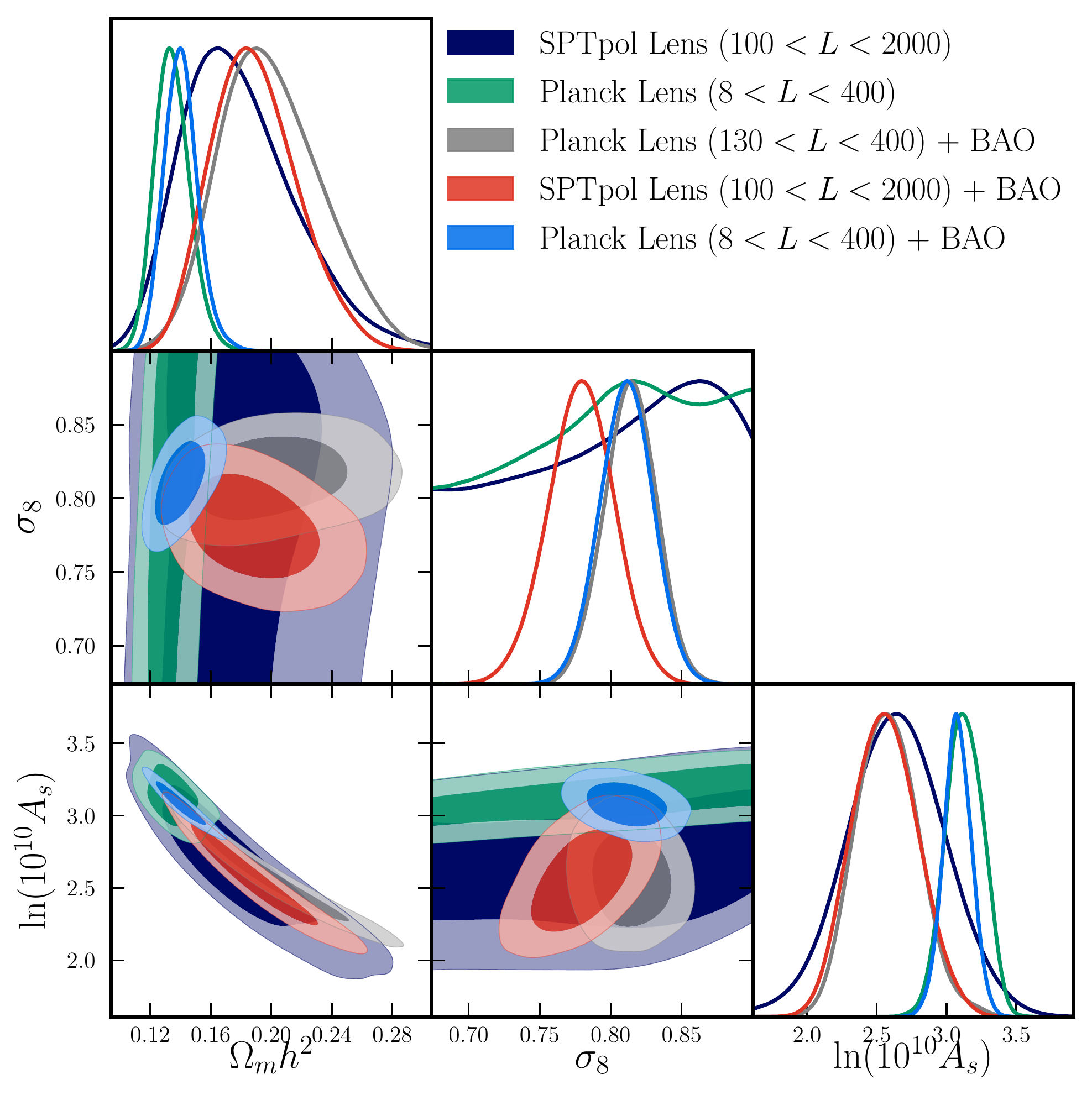}
    \caption{Constraints within the \lcdm model on $\Omega_mh^2$, $\sigma_8$, $\ln(10^{10}A_s)$ from CMB lensing alone and in combination with BAO measurements. We also show the effect of discarding the information about the peak of the CMB lensing power spectrum on the parameter degeneracies (see \planck lensing $130 < L < 400$ + BAO). Contours contain 68\% and 95\% of the posteriors.}
    \label{fig:triangle_plot}
\end{figure}

Combining \sptpol\ lensing with BAO, we obtain the following \lcdm constraints (68\%):

\beq
\begin{rcases}
  H_0 &= 72.0^{+2.1}_{-2.5}\,\text{km}\,\text{s}^{-1}\,\text{Mpc}^{-1} \\
  \sigma_8 &= \sigmaSPTpolLensBAO \\
  \Omega_m &= \OmegamSPTpolLensBAO
\end{rcases}
\text{\sptpol\ lensing + BAO,}
\eeq
while combining \planck lensing with BAO yields:
\beq
\begin{rcases}
  H_0 &= 67.9^{+1.1}_{-1.3}\,\text{km}\,\text{s}^{-1}\,\text{Mpc}^{-1} \\
  \sigma_8 &= 0.811 \pm 0.019 \\
  \Omega_m &= 0.303^{+0.016}_{-0.018}
\end{rcases}
\text{\planck lensing + BAO.}
\eeq
A summary of the constraints from CMB lensing, both \planck and \sptpol{} datasets, and BAO data is provided in Tab.~\ref{tab:cmblens_bao}.
In Fig.~\ref{fig:triangle_plot} we show the constraints on $\Omega_mh^2$, $\sigma_8$, and $\ln(10^{10}A_s)$ obtained with \sptpol{} lensing + BAO (red contours) and \planck lensing + BAO (blue contours).
The \sptpol\ + BAO set prefers higher $H_0$ and $\Omega_m$ than the \planck+ BAO set. Since the included BAO measurements are the same for both cases, 
the differences in the best-fit parameters are indicative of the different
preferences of the two sets of lensing band powers.

To understand the parameter preferences from the two experiments, we first note that $H_0$ and $\Omega_m$ correlate positively in the posterior distribution of the BAO measurements with priors on $\Omega_bh^2$~\citep[e.g.,][]{addison18}. 
As discussed in \citet{planck15-15}, the shape of the \planck\ lensing spectrum constrains $\Omega_m^{0.6} h \approx$ constant, preferring an anti-correlation between $H_0$ and $\Omega_m$ and thus breaking degeneracies of these parameters.
The BAO+$\Omega_bh^2$ constraints dominate the $H_0$-$\Omega_m$ degeneracy direction when combined with \sptpol\ CMB lensing. Therefore the preference for higher $H_0$ from \sptpol\ lensing when combined with BAO is driven by the \sptpol\ lensing $H_0$-$\Omega_m$ contours intersecting the BAO $H_0$-$\Omega_m$ contours around larger values of $H_0$ and $\Omega_m$ compared to \planck\ lensing. 
Compared to the \planck\ lensing measurement, \sptpol\ lensing does not measure the peak of the lensing spectrum.
The peak of the lensing spectrum is sensitive to the scale of matter-radiation equality and effectively constrains the matter density $\Omega_mh^2$. Without measurements of the peak, the \sptpol\ lensing measurement allows for a broader degeneracy between $\Omega_mh^2$ and $A_s$.
Indeed, the best-fit $\Omega_mh^2$ and $\ln (10^{10}A_s)$ from the \sptpol\ lensing measurement are $\sim1\sigma$ higher and $\sim1.4\sigma$ lower compared to the best fits of \planck's lensing measurements. 
With this preference for a high $\Omega_mh^2$ from the \sptpol\ lensing spectrum, the constraints on $H_0$ and $\Omega_m$ when combined with BAO are driven high compared to \planck\ lensing. 

To  confirm this intuition, we rerun the \planck lensing + BAO chain discarding the first three band powers covering the peak of the lensing power spectrum. This leaves us with six band powers between $135 < L < 400$ for a naive $S/N \sim \sqrt{\sum_L (C_L^{\phi\phi}/\Delta C_L^{\phi\phi})^2} \sim 25$ (for comparison, \sptpol{} gives us $S/N \sim 18$). As expected, we find that removing the information about the peak of $C_L^{\phi\phi}$ results in a broadening of the $A_s - \Omega_mh^2$ degeneracy and the contours overlap with the \sptpol{} ones, see Fig.~\ref{fig:triangle_plot}. 
Specifically, we find the following constraints

\begin{equation}
\left.
\begin{array}{l}{H_0 = 72.6^{+2.3}_{-2.9}  \text{ km s$^{-1}$ Mpc$^{-1}$}} \\ 
{\sigma_{8}\,\,= 0.814 \pm 0.019} \\ {\Omega_{m}=0.379^{+0.036}_{-0.042}}
\end{array}\right\} 
\begin{array}{l}{\text {\planck lensing }} 
\\ {\text{$135 < L < 400$ + BAO }},
\end{array}
\end{equation}
in agreement with \sptpol{} lensing + BAO. 
Recall that our \sptpol{} CMB lensing + BAO constraints are more sensitive to the low-redshift ($z \lesssim 4$) universe compared to the primary CMB. It is then interesting to compare the constraints on $H_0$ from CMB lensing + BAO to values inferred from the primary CMB and to direct measurements $H_0$.

Given the degeneracy between the $H_0$ and  $\Omega_m$ from the BAO data and the preference for high $\Omega_m$ of \sptpol{} lensing data, the best-fit $H_0$ from \sptpol{} lensing + BAO is  $72.0^{+ 2.1}_{- 2.5}$ km s$^{-1}$ Mpc$^{-1}$. This sits between the supernovae and strong gravitational lensing time-delay $H_0$ values from SH0ES/H0LiCOW \citep{riess19,wong19} and the supernovae based CCHP $H_0$ values \citep{freedman19}, and is within $\sim 1\,\sigma$ of both measurements. 
Compared to that inferred from \planck's primary CMB spectra \citep[$67.27 \pm 0.60$,][\textsc{TT+TE+EE+lowE}]{planck18-6}, the \sptpol{} lensing + BAO $H_0$ value is also $\sim 2\,\sigma$ high.\footnote{We also note that, thanks to the $H_0 - \Omega_m$ degeneracy in the CMB lensing + BAO data case, the matter density value $\Omega_m$ inferred from \sptpol{} lensing + BAO data is similarly larger, at the $\sim 1.5\, \sigma$ level, than the one suggested by primary CMB \citep[$0.3166  \pm 0.0084$,][\textsc{TT+TE+EE+lowE}]{planck18-6}.}
Note however, the $H_0$ value depends on the $L$ range of the data, as discussed earlier. This CMB lensing measurement, when combined with BAO + $\Omega_bh^2$ prior, provides a separate inference on $H_0$ utilizing information from the low-redshift universe.

Let us now look at the constraint on the $\sigma_8$ parameter. 
The value of the amplitude of matter fluctuations suggested by \sptpol{} + BAO is $\sigma_8=\sigmaSPTpolLensBAO$. This is  consistent, at the 1.1\,$\sigma$ level, with the full-sky \planck lensing result of $\sigma_8= 0.811 \pm 0.019$ and 1.8\,$\sigma$ lower than the primary CMB result of $\sigma_8=0.829\pm0.015$. The \sptpol{} + BAO preference for a lower $\sigma_8$ simply represents another way of stating the preference for a lower $A_s$. Interestingly, as also alluded in Sec.~\ref{sec:galaxy_lensing}, there are indications that the $\sigma_8$ value inferred from LSS probes such as clusters \citep[e.g.,][]{dehaan16,bocquet19}, cosmic shear \citep[e.g.,][]{hildebrandt17,joudaki17a,joudaki17b,planck18-8,abbott18,hikage19}, and redshift space distortions \citep[e.g.,][]{gil-marin17}, are lower than what \planck would suggest, although the difference is not as significant as the $H_0$ tension. 

\begin{table}[]
\centering
\caption{Constraints on a subset of \lcdm parameters using the \planck and \sptpol\ CMB lensing datasets alone, jointly analyzed, or combined with BAO information.  All limits in this table are 68\% intervals,  $H_0$ is in units of km s$^{-1}$Mpc$^{-1}$.}
\label{tab:cmblens_bao}
\begin{tabular}{cc|ccc}
\hline 
\hline     
 & \multicolumn{1}{c|}{Lensing} & \multicolumn{3}{c|}{Lensing + BAO} \\ 
  \hline
 & $\sigma_8 \Omega_m^{0.25}$ & $\sigma_8$   & $H_0$ & $\Omega_m$                 \\ 
 \hline
\sptpol          & $0.593 \pm 0.025$          & $0.779 \pm 0.023$ & $72.0^{+2.1}_{-2.5}$ & $0.368^{+0.032}_{-0.037}$      \\ 
\planck          & $0.590 \pm 0.020$                   & $0.811 \pm 0.019$ & $67.9^{+1.1}_{-1.3}$ & $0.303^{+0.016}_{-0.018}$ \\
\hline
\end{tabular}
\end{table}

\subsection{Joint constraints from primary CMB power spectrum and lensing}
Adding primary CMB anisotropy information constrains the angular acoustic scale $\theta_*$ to high precision and, in turn, breaks the degeneracy between $\Omega_m$, $\sigma_8$, and $H_0$ that affects the CMB lensing-only constraints. Conversely, CMB lensing data can improve constraints on the amplitude parameters, for example by breaking the $A_s e^{-2\tau}$ degeneracy through lensing smoothing effects, and on those limited by geometrical degeneracies when measured from primary CMB alone.

\begin{figure}[t]
	\includegraphics[width=\columnwidth]{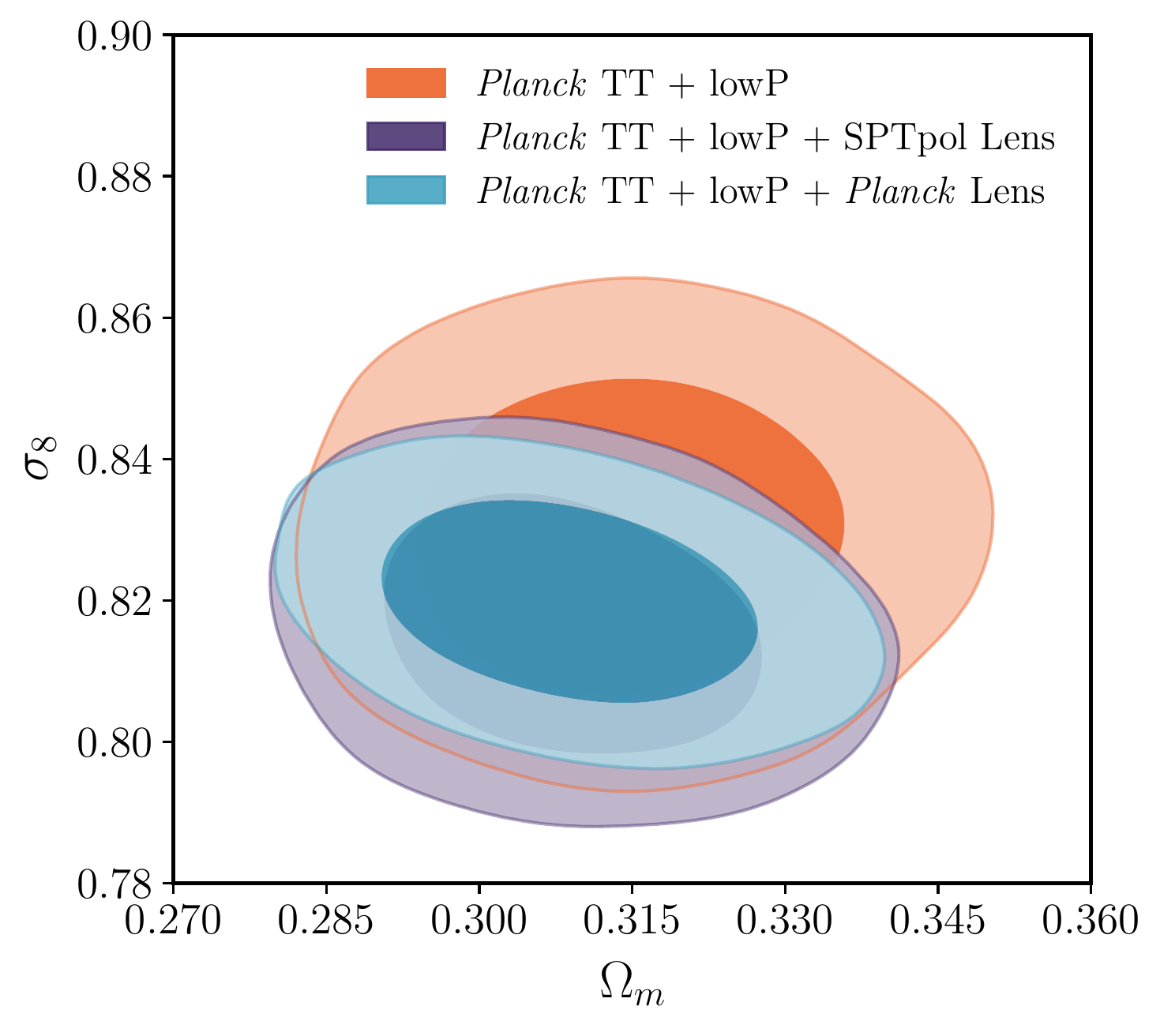}
    \caption{Constraints on $\Omega_m$ and $\sigma_8$ in the base \lcdm model from \planck primary CMB alone (orange contours) and in combination with \planck and \sptpol{} CMB lensing data (blue and purple contours respectively). Contours contain 68 \% and 95 \% of the posteriors.}
    \label{fig:om_sigma8_cmb_cmblens}
\end{figure}

The joint constraints on $\Omega_m$ and $\sigma_8$ from the combination of \planck primary CMB and CMB lensing data are shown in Fig.~\ref{fig:om_sigma8_cmb_cmblens}. Note that in this case, we use the priors shown in the right column of Tab.~\ref{tab:priors} and apply both the response function and $N^{(1)}_L$ linear corrections to the lensing likelihood, as discussed in Sec.~\ref{sec:lensing_like}. 
When primary CMB data are included, the lensing power spectrum shape is almost fixed, but the amplitude still has freedom to increase (decrease) because matter density is allowed to increase (decrease) through the acoustic-scale degeneracy in the primary CMB.
CMB lensing data, either from \sptpol{} or \planck, tend to pull down the $\sigma_8$ value inferred from \planck primary CMB, as also hinted by Fig.~\ref{fig:sigma8_omegam_lensonly}. In particular, we find $\sigma_8 = \sigmaPlanck$ from primary CMB alone (\planckTT + \lowP, 68\%) and $\sigma_8 = \sigmaSPTpolLens$ (\planckTT + \lowP + \sptpol\ lensing, 68\%) and $\sigma_8 = \sigmaPlanckLens$ (\planckTT + \lowP + \planck lensing, 68\%). 
For completeness, we note that that the $\sigma_8$ value inferred from \planck 2018 primary CMB data alone reduces to $\sigma_8 = 0.812 \pm 0.009$ \citep[2018 \textsc{Planck TT+lowE},][]{planck18-6}, thanks to a more precise and lower value of the optical depth, as mentioned in Sec.~\ref{sec:cosmo_inference_framework}.
Finally, we note that the geometrical information from BAO further improves constraints on $\Omega_m$ by roughly 40\% for all datasets considered.

\subsection{Lensing amplitudes}
\label{sec:constraints_lensing_amps}
Gravitational lensing is responsible for transferring CMB power from large to small scales and for smearing the acoustic peaks. Both these effects have been accurately observed and measured in CMB power spectra \citep[e.g.,][]{reichardt09a,story13,louis17,henning18,planck18-6}.

A well-known internal tension in \planck is the preference (at $\approx 2.5\sigma$ significance) of a slightly larger amount of lensing as measured from the smoothing 
of the acoustic peaks, than what is predicted given \lcdm \citep{planck15-13,planck18-6}. At the same time, we note that similar analyses of the \spt\ CMB temperature \citep{story13,aylor17} and polarization \citep{henning18} spectra have found no evidence of this enhanced peak smoothing effect, reporting a mild preference ($\approx 1 \sigma$) for a lower lensing power than predicted.\footnote{The significance of the \spt\ data preference for low lensing power when compared to \planck can be exacerbated if super-sample and intra-sample lensing covariances are neglected, see \citet{motloch19}. }  

The CMB lensing measurement from W19 represents an independent cross-check on the \planck lensing amplitude measurement.
To this end, we follow \citet{planck18-8,simard18} and introduce 
two phenomenologically motivated lensing amplitude parameters, $A_L$ and $A_L^{\phi\phi}$. The former is an unphysical parameter that scales the lensing power spectrum both in the acoustic 
peak smearing and the lens reconstruction, while the latter only scales the theory lens reconstruction at every point in the parameter space.

Marginalizing over $A_L$ effectively removes the lensing information from extra peak-smoothing beyond \lcdm in the \planckTT 2-point function. Then, when both parameters are allowed to vary, the combination $A_L \times A_L^{\phi\phi}$ quantifies the overall amplitude of the measured lensing power with respect to \lcdm expectations, when the inferred \lcdm parameters are insensitive to the observed level of peak smearing.

We start the comparison between \planck and \sptpol{} lensing by fixing $A_L$ to unity. 
The preference for $A_L > 1$ in \planck temperature data pulls the cosmological parameters to a region of the parameter space with a higher intrinsic  CMB lensing power spectrum.
This, in turns, leads the inferred lensing amplitude $A_L^{\phi\phi}$ to lower values. Specifically, we find:
\bea
\label{eq:A_phi_phi_sptpol}
A_L^{\phi\phi} &= \AphiphiSPTpol \quad \text{\sptpol\ Lensing, 68\%},\\
\label{eq:A_phi_phi_planck}
A_L^{\phi\phi} &= \AphiphiPlanck \quad \text{\planck Lensing, 68\%},
\eea 
both in combination with \planckTT and \lowP. 
As can be seen, \planck lensing is consistent within 1$\sigma$ to the \lcdm expectations based on \planck primary CMB, while the \sptpol{} measurement is about 1.8$\sigma$ lower than $A_L^{\phi\phi}=1$.
Note that the \sptpol{}-based $A_L^{\phi\phi}$ value quoted here (Eq.~\ref{eq:A_phi_phi_sptpol}) differs from $A_L^{\phi\phi} = 0.944 \pm 0.058$ (stat) reported in W19 because here we marginalize over the six \lcdm cosmological parameters. An indication of a mild lensing power deficit was also seen in the \sptsz + \planck lensing measurement from \citet{omori17}, for which \citet{simard18} estimate $A_L^{\phi\phi} = 0.91 \pm 0.06$, consistent with both the \sptpol\ and \planck values presented here. 

An informative check to perform is replacing \planckTT with the \sptpol{} {\sc TEEE} dataset. This way we can test the impact of primary CMB on the inferred lensing power spectrum amplitude. 
As expected, the mild \sptpol{} preference for less lensing smoothing pushes the inferred 4-point lensing amplitude to values $\approx1\sigma$ above unity:   

\beq
A_L^{\phi\phi} = 1.13^{+0.13}_{-0.11} \quad \text{\sptpol\ Lensing, 68\%}.
\eeq 
The constraints on $A_L^{\phi\phi}$ from different combinations of datasets are reported in Tab.~\ref{tab:constraints}.

\begin{figure}
	\includegraphics[width=\columnwidth]{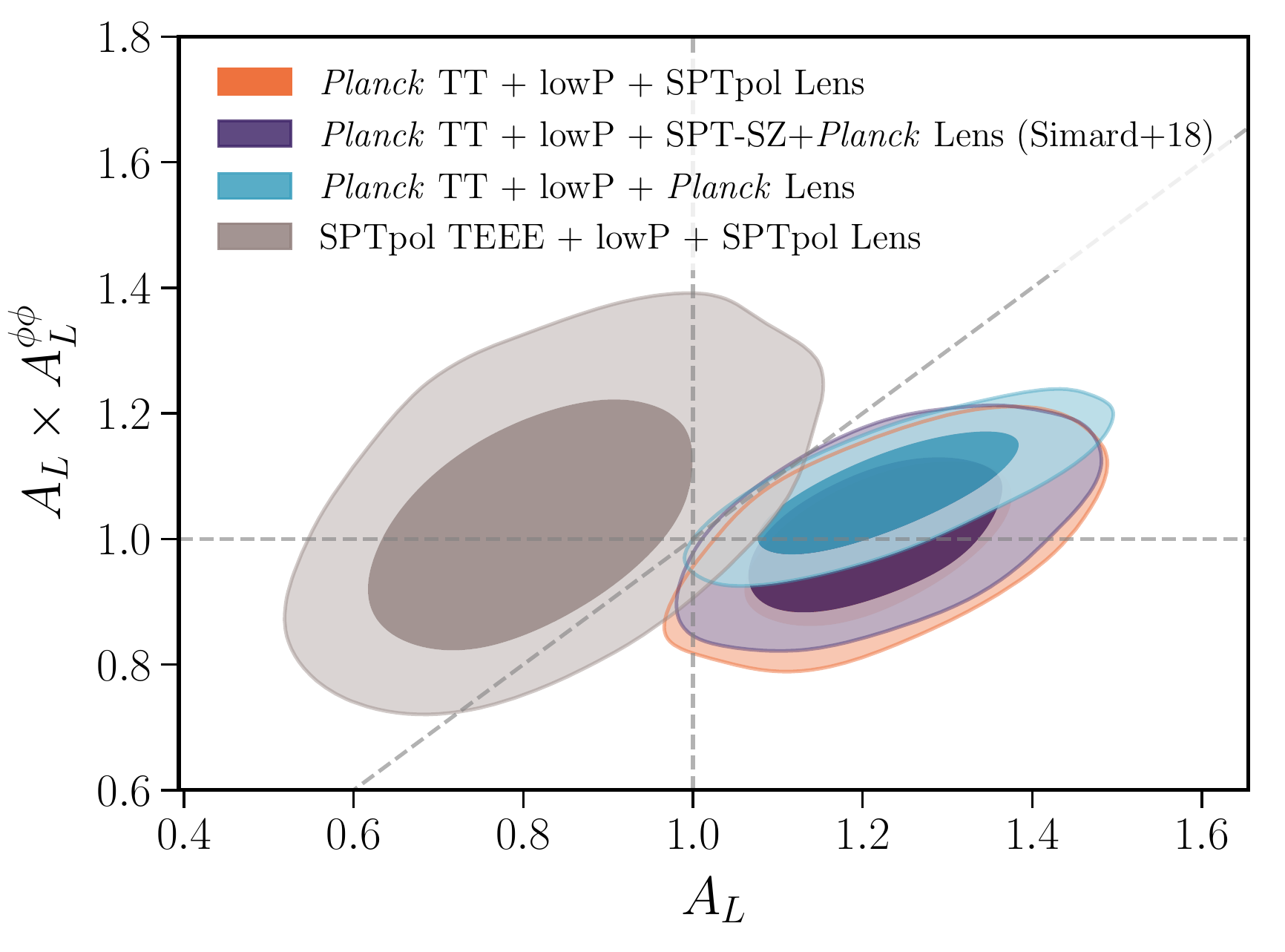}
    \caption{The CMB lensing and primary CMB power spectra are sensitive to the lensing effects in different ways. The acoustic peak smoothing induced by lensing on \planck primary CMB favours models with $A_L > 1$. When the peak smearing information has been marginalized over, the amplitude of the lensing trispectrum relative to the best-fit \lcdm parameters is consistent with expectations independent of the dataset combination. The results based on the \sptsz + \planck lensing map from \citet{omori17} presented in \citet{simard18} are also consistent with the \sptpol\ constraints.}
    \label{fig:lensing_amps}
\end{figure}

The next question we would like to answer is whether the lensing power observed in the 4-point function is consistent 
with \lcdm expectations when the peak smoothing effect, from either \planck or \sptpol{} primary CMB, is not reflected on the cosmological constraints.
To investigate this aspect, we show in Fig.~\ref{fig:lensing_amps} the posterior distributions on $A_L$ and $A_L \times A_L^{\phi\phi}$ using \planck primary CMB  in combination with \planck and \sptpol{} lensing datasets (purple and orange contours respectively). 
When letting both $A_L$ and $A_L^{\phi\phi}$ free to vary, we obtain (with \planckTT and \lowP, 68\%)

\bea
A_L \times A_L^{\phi\phi} &= \AlensAphiphiSPTpol \quad \text{\sptpol\ Lensing,}\\
A_L \times A_L^{\phi\phi} &= \AlensAphiphiPlanck \quad \text{\planck Lensing},
\eea 
These values show that both lensing datasets appear consistent with the cosmological parameters implied by the 2-point function, once peak-smearing effects are marginalized over.
Finally, the \sptpol{} lensing dataset is also consistent with \lcdm expectations when \planck primary CMB is replaced with \sptpol{} $TEEE$, with information from peak smoothing marginalized over:

\beq
A_L \times A_L^{\phi\phi} = \AlensAphiphiSPTpolSPTpol \quad \text{\sptpol\ Lensing, 68\%}.
\eeq 
The individual constraints on $A_L$ and $A_L^{\phi\phi}$ when both are allowed to vary are summarized in Tab.~\ref{tab:constraints}. Note how both \planck and \sptpol{} preferences for $A_L \neq 1$ are preserved even when $A_L^{\phi\phi}$ is included as an additional parameter. This demonstrates that the driver of $A_L$ best-fit values is the features in both the \planck and \sptpol{} 2-point CMB spectra.

\subsection{Massive neutrinos}
\label{sec:constraints_neutrinos}
We now turn to examine what CMB lensing measurements tell us about fundamental physics, specifically about neutrino properties. Despite the fact that neutrino oscillation measurements have established that neutrinos are massive, their absolute mass scale and the relative ordering of the mass eigenstates - the so-called neutrino hierarchy - are still largely unknown. Neutrino oscillation experiments are sensitive to the squared mass differences and suggest that the sum of the neutrino masses is $\sum m_{\nu} > $ 58 meV in the normal hierarchy and  $>$ 100 meV in the inverted hierarchy \citep[][and references therein]{desalas17}. 
Interestingly, the current generation of long baseline neutrino oscillation experiments such as T2K\footnote{\url{https://t2k-experiment.org/}} and NO$\nu$A\footnote{\url{https://novaexperiment.fnal.gov/}}, which are mostly sensitive to the mass hierarchy, have found a mild preference for the normal hierarchy \citep{abe2017,acero2019}.

In the context of neutrino studies, cosmological observations greatly complement laboratory measurements as they enable a constraint of the sum of the neutrino masses \citep[e.g.,][]{vagnozzi17}. 
In particular, the CMB lensing potential power spectrum is sensitive to $\sum m_{\nu}$ since massive neutrinos suppress the growth of structure below the neutrino free-streaming length, resulting in a scale-dependent suppression of $C_L^{\phi\phi}$.

\begin{figure}
	\includegraphics[width=\columnwidth]{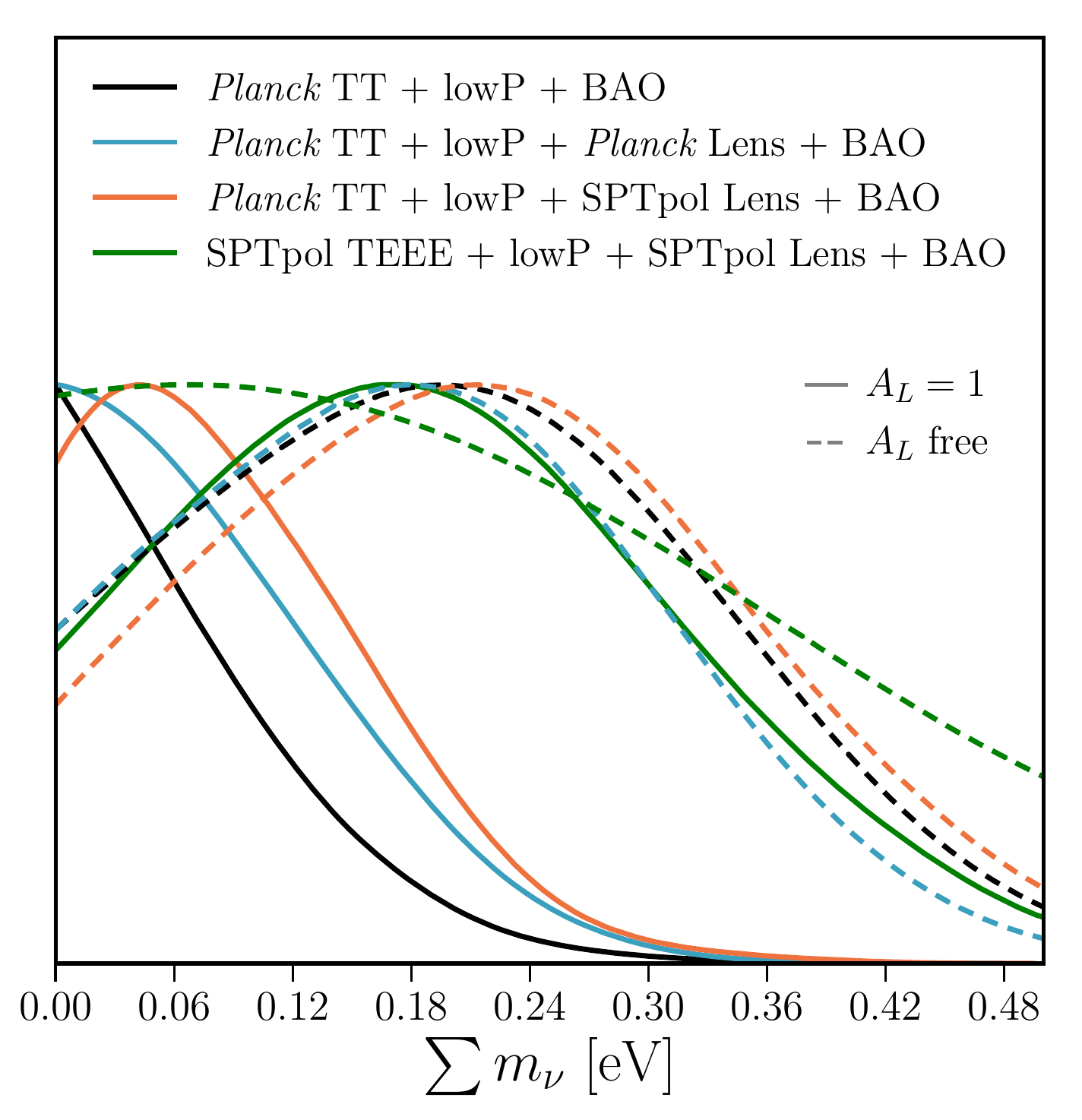}
    \caption{Constraints on the sum of the neutrino masses $\sum m_{\nu}$ when \planck primary CMB and BAO information is exploited (black line) and when either \sptpol\ lensing (orange line) or \planck lensing (cyan line) is included in the cosmological analysis. If we replace the \planck primary CMB with the \sptpol{} {\sc TEEE} measurement from \citet{henning18} and include \sptpol{} lensing and BAO we obtain the green curve. Dashed lines show instead the results when we marginalize over the lensing information in the primary CMB, i.e. we let $A_L$ free to vary.}
    \label{fig:mnu}
\end{figure}

Let us first look at the constraints on $\sum m_{\nu}$ from primary CMB alone. \planck constrains the sum of the neutrino masses to $\sum m_{\nu} < 0.69$ eV at 95\% level (\planckTT + \lowP).
This upper limit can be further improved by adding data on the BAO scale, as the low-redshift information allows us to break parameter degeneracies, for instance between $\sum m_{\nu}$ and $H_0$.
With this setup, we obtain $\sum m_{\nu} < 0.20$ eV (95\%), which is shown by the black solid line in Fig.~\ref{fig:mnu} (for the remainder of this subsection we always include    
BAO data unless otherwise stated). 
As mentioned in Sec.~\ref{sec:constraints_lensing_amps}, the amount of lensing inferred from primary CMB is larger than the one directly measured through the amplitude of the lensing power spectrum. 
Therefore, the constraints on $\sum m_{\nu}$ from primary CMB alone (+BAO) are tighter when CMB lensing is not included.
This is because increasing the neutrino mass corresponds to a decrease in the acoustic peak smearing expected within \lcdm.

In fact, after the inclusion of CMB lensing information we obtain:

\beq
\begin{split}
\sum m_{\nu} < \mnuSPTpolLensBAO\, \text{eV}  \quad (&\text{\planckTT + \lowP+ BAO}\\
& \text{+\sptpol\ lensing},\,\, 95\%),
\end{split}
\eeq
\beq
\begin{split}
\sum m_{\nu} < \mnuPlanckLensBAO\, \text{eV} \ \quad (&\text{\planckTT + \lowP+ BAO}\\
& \text{+\planck lensing},\,\, 95\%),
\end{split}
\eeq
shown as the solid orange and cyan lines in Fig.~\ref{fig:mnu}. 
These results are in good agreement with each other. For a direct comparison with the previous \sptsz lensing measurement, \citet{simard18} find a 95\% upper limit on the sum of the neutrino masses of $\sum m_{\nu} < 0.70$ eV, while we obtain  $\sum m_{\nu} < 0.72$ eV when replacing \sptsz+\planck lensing with \sptpol\ lensing, both without BAO. 

An instructive test to check the stability of the neutrino mass constraints with respect to changes in the primary CMB is replacing the \planckTT likelihood with the \sptpol{} {\sc TEEE} one  from \citet{henning18}. This test is especially interesting because, as we have already mentioned, the \planck and \sptpol{} primary CMB measurements are known to favour different amount of lensing from the smoothing of the acoustic peaks.
In fact, this dataset combination (the solid green line in Fig.~\ref{fig:mnu}) suggests

\beq
\begin{split}
\sum m_{\nu} < \mnuSPTpolTEEESPTpolLensBAO\, \text{eV} \ \quad (&\text{\sptpol{} TEEE + \lowP+ BAO}\\
& \text{+\sptpol{} lensing},\,\, 95\%),
\end{split}
\eeq
which is larger than what is found using the temperature and large scale polarization from \planck. Differently from \planckTT measurement, the high-$\ell$ \sptpol{} {\sc TEEE} spectra prefer slightly less lensing than in base \lcdm ($1.4\,\sigma$ below $A_L = 1.0$ and $2.9\,\sigma$ lower than the value preferred by \planckTT). In turn, the lensing power deficit is interpreted as a larger neutrino mass due to their structure suppression effect, pushing the constraints on $\sum m_{\nu}$ to larger values.

Finally, we free the lensing amplitude parameter $A_L$ and investigate the \sptpol\ lensing constraining power when we marginalize over the effect on $\sum m_{\nu}$ from excess peak smoothing of the primary \planck 2-point measurements. This is particularly interesting in light of \planck's $A_L$ being 2.5\,$\sigma$ high compared to \LCDM\ expectation \citep{planck18-6}. The results are shown as the dashed lines in Fig.~\ref{fig:mnu} with the same color coding introduced above.

When using \planckTT as the primary CMB, $A_L$ takes on values greater than 1 due to the significant constraining power of the 2-point power spectrum. This, compared to when $A_L$ is fixed to 1, lets the matter parameters take on lower values and allows for a larger value of $\sum m_{\nu}$. As a result, the 95\% C.L. upper limits on $\sum m_{\nu}$ from both \planck lensing and \sptpol\ lensing increase to:
\beq
\begin{split}
\sum m_{\nu} < \mnuSPTpolLensBAOAlens\, \text{eV}  \quad (&\text{\planckTT + \lowP+ BAO}\\
& \text{+\sptpol\ lensing  [$A_L$ free]}, \,\,95\%),
\end{split}
\eeq
\beq
\begin{split}
\sum m_{\nu} < \mnuPlanckLensBAOAlens\, \text{eV} \ \quad (&\text{\planckTT + \lowP+ BAO}\\
& \text{+\planck lensing [$A_L$ free]}, \,\,95\%).
\end{split}
\eeq

As can be noted from Tab.~\ref{tab:constraints}, when CMB lensing likelihood is included in the cosmological inference, the central value of $A_L$ is still larger than unity (e.g., $A_L=1.15^{+0.09}_{-0.12}$ for \sptpol\ lensing), while for primary CMB + BAO we find  $A_L=1.28^{+0.10}_{-0.13}$. 

On the other hand, when using the \sptpol{} $TEEE$ measurement instead of \planckTT, $A_L$, instead of having a 1.4\,$\sigma$ lower value as would be preferred by \sptpol\ TEEE, takes on the value $A_L=1.03^{+0.28}_{-0.23}$. Thus the best-fit posterior values in this $A_L$-free chain are similar to the $A_L = 1$ chain. However, since $\sum m_{\nu}$ and $A_L$ are degenerate (positively correlated), including $A_L$ as a free parameter essentially broadens the posterior distribution of $\sum m_{\nu}$. Therefore, we obtain a larger upper limit on $\sum m_{\nu}$, specifically we find:

\beq
\begin{split}
\sum m_{\nu} < \mnuSPTpolTEEESPTpolLensBAOAlens\, \text{eV} \ \quad (&\text{\sptpol{} TEEE + \lowP+ BAO}\\
& \text{+\sptpol{} lensing [$A_L$ free]},\,\, 95\%).
\end{split}
\eeq
The constraint on the sum of neutrino masses from \sptpol\ lensing is consistent with \planck lensing, allowing slightly higher neutrino mass because of the overall smaller lensing amplitude.

\begin{table*}[]
\centering
\caption{Constraints on several extensions to the base six parameters \lcdm model for combinations of primary CMB and lensing power spectra from \planck and \sptpol{}. Horizontal lines separate the different cosmological models that have been analyzed. All limits are 68\% except on $\sum m_{\nu}$, for which we report the 95\% upper limits. Note that the results for cosmological runs with varying $\sum m_{\nu}$ also include BAO information in addition to the datasets shown in the first row. The number in parenthesis in the $\Omega_K$ run also shows the effect of the BAO data inclusion.}
\label{tab:constraints}
\resizebox{\textwidth}{!}{%
\begin{tabular}{ccccc}
\hline
\hline
\multirow{2}{*}{}           & \multirow{2}{*}{TT + lowP} & \multirow{2}{*}{\begin{tabular}[c]{@{}c@{}}TT + lowP + \\ \sptpol\ Lens\end{tabular}} & \multirow{2}{*}{\begin{tabular}[c]{@{}c@{}}TT + lowP + \\  \planck Lens\end{tabular}} & \multirow{2}{*}{\begin{tabular}[c]{@{}c@{}}\sptpol\ TEEE + lowP + \\ \sptpol\ Lens\end{tabular}} \\
                            &                            &                                                                                     &                                                                                      &                                                                                              \\ 
\hline
\hline
$A_L^{\phi\phi}$            & $\dots$                    & $0.890^{+0.057}_{-0.066}$                                                           & $0.970 \pm 0.039$                                                                    & $1.13^{+0.11}_{-0.13}$                                                                       \\
\hline
$A_L^{\phi\phi}$            & $\dots$                    & $0.817 \pm 0.065$                                                                   & $0.876^{+0.042}_{-0.052}$                                                            & $1.27^{+0.15}_{-0.21}$                                                                       \\
$A_L$                       & $\dots$                    & $1.222^{+0.097}_{-0.11}$                                                            & $1.233^{+0.093}_{-0.11}$                                                             & $0.70^{+0.15}_{-0.20}$                                                                       \\
$A_L \times A_L^{\phi\phi}$ & $\dots$                    & $0.995 \pm 0.090$                                                                   & $1.076 \pm 0.063$                                                                    & $1.036 \pm 0.136$                                                                            \\
\hline
$\sum m_{\nu}$    [eV]          & \textless 0.196            & \textless  0.229                                                                     & \textless 0.223                                                                       & \textless 0.420                                                                              \\
\hline
$\sum m_{\nu}$ [eV]              & \textless 0.430            & \textless 0.453                                                                     & \textless 0.394                                                                      & \textless 0.620                                                                              \\
$A_L$                       & $1.28^{+0.10}_{-0.13}$     & $1.15^{+0.09}_{-0.12}$                                                              & $1.11^{+0.07}_{-0.08}$                                                               & $1.03^{+0.09}_{-0.15}$                                                                       \\
\hline
\multirow{2}{*}{$\Omega_K$} & $-0.050^{+0.030}_{-0.017}$ & $-0.0099^{+0.013}_{-0.0084}$                                                        & $-0.0084^{+0.0093}_{-0.0076}$                                                        & $\dots$                                                                                      \\
                            & ($0.0005 \pm 0.0026$)      & $(0.0007 \pm 0.0025$)                                                               & ($0.0002 \pm 0.0026$)                                                                & $\dots$                 \\
\hline
\hline                                                           
\end{tabular}%
}
\end{table*}

\subsection{Spatial curvature}
\label{sec:constraints_curvature}
A general prediction of inflationary models is the flatness of the spatial hyper-surfaces of the background metric.
A main hindrance in determining the geometry of the universe solely from primary CMB observations is the well-known geometrical degeneracy \citep{efstathiou99}.
This degeneracy arises because the shape of the CMB anisotropy spectrum mainly depends on two physical scales, the sound horizon at recombination and the angular diameter 
distance to the last scattering surface, so that cosmological models with similar matter content and angular diameter distance to the last scattering surface  will produce nearly 
indistinguishable CMB power spectra.

\begin{figure}
	\includegraphics[width=\columnwidth]{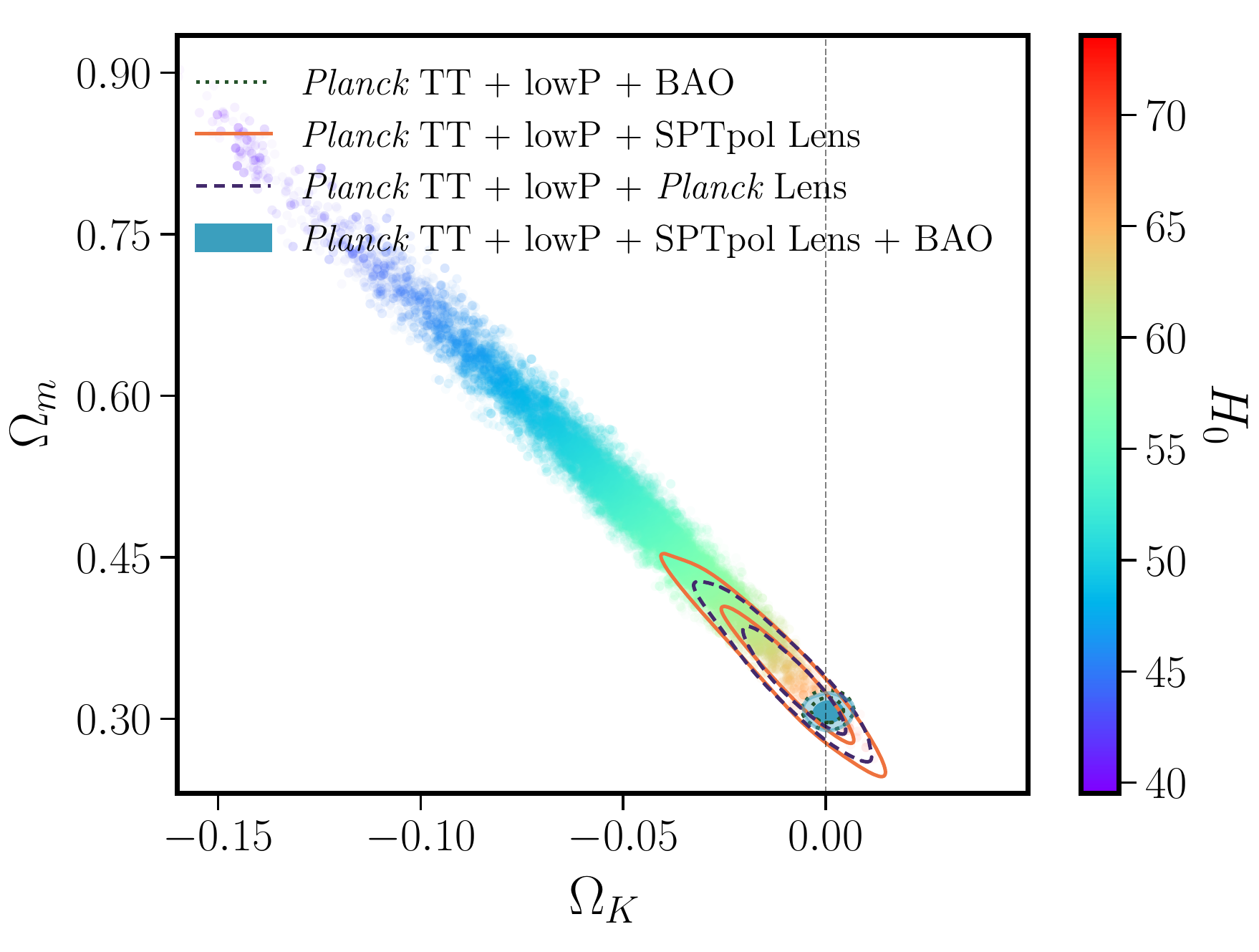}
    \caption{Constraints on curvature and total matter density from primary CMB (scattered points color-coded by Hubble constant value). Closed universe models with high curvature are inconsistent with lensing measurements (solid red and dashed purple lines, \sptpol\ and \planck lensing respectively) and ruled out by BAO data (dotted green line).  The joint analysis of \planck primary CMB, \sptpol\ lensing, and BAO is fully consistent with a flat universe (blue shaded contour).}
    \label{fig:omega_K}
\end{figure}

The geometrical degeneracy is manifest when looking at the coloured scattered points in Fig.~\ref{fig:omega_K} that have been obtained using only \planck primary CMB data. In particular, the \planckTT preference for larger $A_L$ values allows the degeneracy to extend to regions of the parameter space with low Hubble constant and negative curvature.
This picture can be greatly improved by using either internal CMB data alone, specifically by adding measurements of CMB lensing that break the geometrical degeneracy, or through the inclusion of BAO data. 
The constraint on spatial curvature from \planck primary CMB only is $\Omega_K = \omegaKPlanck$, favouring a positive curvature at about $1.5\,\sigma$. Instead, the inclusion of \sptpol\ CMB lensing yields

\beq
\begin{split}
\Omega_K = \omegaKSPTpolLens  \quad (&\text{\planckTT + \lowP}\\
& \text{+\sptpol\ lensing}, 68\%)\\
&
\end{split}
\eeq
in agreement with the \planck lensing based result of $\Omega_K = \omegaKPlanckLens$. 

Finally, external data like BAO also provide consistent results when combined with primary CMB \planck data (see Fig.~\ref{fig:omega_K}). By jointly analysing \planckTT + \lowP + \sptpol\ lensing + BAO we find $\Omega_K = \omegaKSPTpolLensBAO$, a sub-percent measurement of the spatial curvature of the universe.

%%%%%%%%
% CONCLUSIONS
%%%%%%%%
\section{Conclusions}
\label{sec:conclusions}
The lensing band powers presented by \citet{wu19} are currently the most precise measurement of the CMB lensing power spectrum from the ground. 
As such, the band powers present a valuable, independent check on the full-sky \planck lensing measurement,  and also extend the measurement to smaller angular scales. 
In this work, we investigate the cosmological implications of these data, and explore the tensions that are emerging between the high- and low-redshift universe within the \lcdm framework. 

Overall, the constraints based on \sptpol{} lensing are in close agreement with the ones obtained on the full-sky with \planck.
For example, using only \sptpol{} CMB lensing data, we find a 4.2\% constraint on $\sigma_8\Omega_m^{0.25} = \sigmaOmSPTpol$, matching the \planck based value 
of  $\sigma_8\Omega_m^{0.25} = \sigmaOmPlanck$. If we further combine the \sptpol{} and \planck lensing likelihoods, we improve the constraint precision from CMB lensing alone to 2.5\%, $\sigma_8\Omega_m^{0.25} = \sigmaOmPlanckSPTpol$.
When complementing the \sptpol{} lensing likelihood with BAO data, the constraints tighten to $\sigma_8=\sigmaSPTpolLensBAO$ and $\Omega_m = \OmegamSPTpolLensBAO$, which when compared to similar constraints using \planck lensing with BAO data are $\sim1.5\,\sigma$ lower and higher, respectively.
We identify the lack of information about the peak of the CMB lensing spectrum from the \sptpol{} data to be the driving factor of this difference.

The \sptpol{} lensing band powers also provide an informative cross-check on the internal \planck tension that exists between the amount of lensing directly measured from the 4-point function reconstruction and the one inferred from the acoustic peak smearing. 
In particular, the lensing amplitude measured from \sptpol{} is consistent (albeit $\approx 1\,\sigma$ low) with the one inferred from the \planck lensing reconstruction, and in tension with that deduced from CMB peak smearing. 
When the sensitivity to lensing is removed from the peak smearing effect in the CMB 2-point function, the \sptpol{} data match the amount of lensing predicted by the observed primary CMB anisotropies.

When combined with \planck primary CMB data, the \sptpol{} lensing and \planck lensing constraints agree. Among the single-parameter extensions to the \lcdm{} model that we consider, the spatial curvature is constrained to be  
$\Omega_K = \omegaKSPTpolLensBAO$, while the sum of the neutrino masses $\sum m_{\nu} < \mnuSPTpolLensBAO$ eV at 
 95\% confidence (both including BAO data). 

The preference for a larger lensing signal in the \planck CMB 2-point function is known to drive tighter constraints on the sum of the neutrino masses \citep{planck18-6}. If we remove the 2-point lensing signal from the \planckTT peak smearing by marginalizing over $A_L$, the constraint on $\sum m_{\nu}$ broadens to $\sum m_{\nu} < \mnuSPTpolLensBAOAlens$ eV.
Conversely, when replacing the \planck primary CMB with the \sptpol{} TEEE band powers from \citet{henning18}, which favour $A_L < 1$, we find  $\sum m_{\nu} < \mnuSPTpolTEEESPTpolLensBAO$ eV (fixing $A_L$ to unity).

The cosmological constraints presented in this paper are also in excellent agreement with those obtained from the \sptsz 
temperature-based lensing reconstruction over 2500\,\sqdeg{} \citep{omori17,simard18}, of which the \sptpol\ footprint is a subset.

CMB lensing measurements are becoming increasingly important to precision tests of cosmology. 
In the upcoming years, high-$S/N$ lensing measurements will provide invaluable insights on the growth of structure and the sum of the neutrino masses. 
Current experiments like SPT-3G \citep{benson14, bender18}, as well as future ground-based observations from Simons Observatory \citep{so18} and CMB-S4 \citep{cmbs4-sb1}, are projected to significantly improve constraints on the sum of neutrino masses through CMB lensing, with CMB-S4 obtaining a sufficient sensitivity ($\sim 20$\, meV\footnote{Note that this forecast includes projected measurements of the BAO scale from DESI redshift survey.}) to detect the minimum mass in the normal hierarchy at a significance of $3\,\sigma$.
Estimating and removing the CMB lensing signal, a process known as delensing \citep[e.g.,][]{smith09,manzotti17},  will also be crucial to searches for primordial gravitational waves from inflation.
The ultra-low-noise maps of 1500\,\sqdeg{} of sky from the on-going SPT-3G survey, the latest instrument on the South Pole Telescope \citep{benson14,bender18}, will dramatically improve the lensing reconstruction across this area and our knowledge of high-redshift structure growth.

%%%%%%%%%%%%%%%%%%%%% Ackn., bib, appendix %%%%%%%%%%%%%%%%%%%%%
\begin{acknowledgements}
SPT is supported by the National Science Foundation through grant PLR-1248097.  Partial support is also provided by the NSF Physics Frontier Center grant PHY-1125897 to the Kavli Institute of Cosmological Physics at the University of Chicago, the Kavli Foundation and the Gordon and Betty Moore Foundation grant GBMF 947. This research used resources of the National Energy Research Scientific Computing Center (NERSC), a DOE Office of Science User Facility supported by the Office of Science of the U.S. Department of Energy under Contract No. DE-AC02-05CH11231.  
The Melbourne group acknowledges support from the University of Melbourne and an Australian Research Council's Future
Fellowship (FT150100074). 
Work at Argonne National Lab is supported by UChicago Argonne LLC, Operator  of  Argonne  National  Laboratory  (Argonne). Argonne, a U.S. Department of Energy Office of Science Laboratory,  is  operated  under  contract  no.   DE-AC02-06CH11357.  We also acknowledge support from the Argonne  Center  for  Nanoscale  Materials. 
 
%We acknowledge the use of many python packages: \sc{IPython} \citep{ipython}, \sc{matplotlib} \citep{Hunter:2007}, and \sc{scipy} \citep{scipy}.}.

\end{acknowledgements}
\bibliographystyle{aasjournal}
\bibliography{spt}

\end{document}